\documentclass[aip,amsmath,amssymb,reprint]{revtex4-1}
\usepackage{graphicx}
\usepackage{dcolumn}
\usepackage{bm}

\usepackage[utf8]{inputenc}
\usepackage[T1]{fontenc}
\usepackage{mathptmx}
\usepackage{etoolbox}
\usepackage{hyperref}
\usepackage[utf8]{inputenc}
\usepackage{newunicodechar}
\newunicodechar{−}{-}

\makeatletter
\def\@email#1#2{%
 \endgroup
 \patchcmd{\titleblock@produce}
  {\frontmatter@RRAPformat}
  {\frontmatter@RRAPformat{\produce@RRAP{*#1\href{mailto:#2}{#2}}}\frontmatter@RRAPformat}
  {}{}
}%
\makeatother
\begin{document}

\preprint{AIP/123-QED}

\title{Unveiling the Face-Dependent Ice Growth Kinetics: Insights from Molecular Dynamics on the Basal and Prism Surfaces}

\author{Jihong Shi}
\affiliation{Department of Physics, King's College London, Strand, London WC2R 2LS, UK.
}
\author{Maxwell Fulford}
\affiliation{Department of Physics, King's College London, Strand, London WC2R 2LS, UK.} 
\author{Matteo Salvalaglio}
\affiliation{Department of Chemical Engineering, University College London, Torrington Place, London WC1E 7JE, UK.}
\author{Carla Molteni$^{\ast}$}
 \email{carla.molteni@kcl.ac.uk}
\affiliation{Department of Physics, King's College London, Strand, London WC2R 2LS, UK.
}

\date{\today}

\begin{abstract}
Ice nucleation and growth are critical in many fields, including atmospheric science, cryobiology, and aviation. However, understanding the detailed mechanisms of ice crystal growth remains challenging. In this work, crystallization at the ice/quasi-liquid layer (QLL) interface of the basal and primary prism (prism1) surfaces of hexagonal ice (Ih) was investigated using molecular dynamics simulations across a wide range of temperatures for the TIP4P/Ice model, with comparisons to the mW coarse-grained model. Together with elucidating the temperature-dependent mechanisms of crystallization, face-specific growth rates were systematically estimated. While the prism surface generally exhibits faster growth rates than the basal surface, a temperature-dependent crossover in growth rates between the basal and prism surfaces is observed in TIP4P/Ice simulations, which correlates with crossovers in QLL thickness and properties and with the well-known column to platelets transition in ice-crystal habits at low vapor pressure. This observation helps decode the complex dependence between crystal morphology and temperature in ice crystals.
\end{abstract}

\maketitle


\section{Introduction}
Understanding ice nucleation and regulating ice growth is important for a range of fields, including cryobiology \cite{SEI2002218,xue2015quantifying}, cryopreservation \cite{doi:10.1021/acs.chemrev.5b00744,doi:10.2989/SF.2009.71.2.12.827}, the aviation industry \cite{CharacterizationsofAircraftIcingEnvironmentsthatIncludeSupercooledLargeDrops,doi:10.1063/1.2979247}, and atmospheric science \cite{ProductionofIceinTroposphericCloudsAReview,doi:10.1126/science.1076964}. Even though the characteristics of ice and water have been known for decades, insights into ice crystal growth remain elusive. 

Ice crystals predominantly grow in two types of environmental conditions: supercooled water and supersaturated water vapor \cite{libbrecht2017physical}. However, the presence of a quasi-liquid layer (QLL)\cite{abraham1981phases}, which develops at ice surfaces in the vapor environment, blurs this distinction, as an interface with the liquid exists at both the ice-QLL-vapor and liquid-ice interfaces. At the liquid-ice interface, water molecules from the liquid phase are incorporated into the crystalline structure of ice, growing the ice crystal. This is a direct phase transition from liquid to solid in the melt phase. Similarly, the presence of the QLL between the vapor and ice phases facilitates ice growth from the vapor phase because it allows for the diffusion of water molecules condensed from the vapor onto the interface with ice \cite{libbrecht2015experimental}, which then integrates them into the crystalline lattice. 

The QLL has been extensively studied, including investigations of its thickness, which increases with temperature (see, e.g., reviews by \citeauthor{slater2019surface}\cite{slater2019surface} and \citeauthor{nagata2019}\cite{nagata2019} and references therein). While the Ih basal face grows from both liquid and vapor, in the first environment, the relevant prismatic face for growth is the secondary one (prism2), while in the second, it is the primary one (prism1)\cite{shultz2018crystal}. 

Ice crystal growth is highly sensitive to temperature, as it is a competitive process between molecular ordering and disordering aided by diffusion \cite{raz2005,rozmanov2011temperature}. The driving forces of supercooling for crystallization become more significant as the temperature gradually decreases from the melting point of ice. On the contrary, lowering the temperature reduces the diffusion capacity of molecules, and weaker liquid mobility would be detrimental to ice growth. Thus, mutually opposite effects act on ice growth as the temperature changes. 

The trigger of this work has been the intriguing temperature-dependent hexagonal prism morphologies of ice crystal at low vapor pressure, growing as plate-like, column-like, plate-like, and/or column-like again as temperature decreases \cite{libbrecht2005physics,libbrecht2015experimental,libbrecht2017physical,bailey2009}. Hence, we concentrate here on the basal and prism1 surfaces and aim to unravel the mechanism and kinetics of crystal growth as a function of temperature at the molecular level using molecular dynamics (MD) simulations.

Our approach is motivated by the fact that, at low supersaturation, the growth process is dominated by surface integration, i.e., the incorporation of water molecules in the crystalline lattice. This assumption is supported by the fact that ice crystals adopt a convex shape at low supersaturation rather than the dendritic growth characteristic of higher vapor pressures that lead to the paradigmatic snowflake morphologies. When growth is dominated by surface integration, the limiting timescale for the growth process is associated with the transition of water molecules from the QLL to the crystal. In contrast, the adsorption of water molecules at the QLL-vapor interface can be considered fast. The simulations performed in this work implement this assumption by introducing a finite but large reservoir of liquid-like molecules in the form of an out-of-equilibrium QLL, which relaxes to equilibrium due to the sole effect of the transition of water molecules from the QLL to the crystal.

Ice growth has been extensively studied both from vapor and from the melt, experimentally and theoretically, and a selection of findings is reported below. Although our focus is on the relevant surfaces for vapor growth, studies of growth for the melt may also be relevant as there are some similarities and differences when surfaces with QLL are considered. 

There have been many important experimental investigations related to the growth of ice from vapor \cite{lamb1972linear,furukawa1987ellipsometric,raymond1989inhibition,furukawa1993temperature} and they are constantly being updated,\cite{inomata2018temperature,sazaki2021situ} with advances and developments in optical microscopy playing a crucial role in furthering our understanding of ice growth kinetics. Recently, for example, \citeauthor{sazaki2021situ} \cite{sazaki2021situ} used advanced optical microscopy techniques, laser confocal microscopy combined with differential interference contrast microscopy (LCM-DIM), and visualized a 2D nucleation and screw dislocations growth mechanisms on the ice Ih basal surface, discussing the temperature dependence of growth kinetics and the role of surface diffusion of water molecules. They demonstrated the significant role of surface diffusion of water molecules on the basal face in the lateral growth of elementary steps when the distance between adjacent spiral steps is smaller than 15 $\mu \mathrm{m}$. \citeauthor{miyamoto2022growth}\cite{miyamoto2022growth} observed the growth of elementary spiral steps on the ice Ih prism faces heteroepitaxially grown from vapor on a CdSe substrate using LCM-DIM. The most significant finding was that screw dislocations on prism faces are primarily located in the interiors of the faces, in contrast with basal faces where dislocations are found at the edges, highlighting significant differences in growth behaviors between the prism and basal faces of ice Ih crystals. In parallel, other experiments focused on ice crystallization in solution, e.g., in the presence of antifreeze proteins, using bright-field microscopy and Mach-Zehnder interferometry\cite{vorontsov2018growth,bayer2018growth}, which enable precise measurements of the ice growth rate from liquid at the nm/s scale and the three-dimensional surface structure of ice crystals during growth. Ice growth from nanofilms of supercooled water was also investigated experimentally with a pulse-laser heating technique\cite{xu2016growth}. 

The modelling of ice crystal growth at the molecular scale has been developing for several years and is still evolving. Some of the latest research used machine learning potentials, aimed to reproduce ab initio accuracy, to study the ice nucleation from bulk liquid water on mineral surfaces\cite{soni2024using,piaggi2024first}, ice seed homogenous growth from water\cite{piaggi2022homogeneous}, and ice growth and kinetics at the ice-liquid interface in the bulk\cite{montero2023kinetics,wang2024five}. However, most previous and current research has been based on empirical force fields, which have been demonstrated to work well and concentrate on growth from water in bulk periodic systems, equivalent to growth from the melt, with less numerous studies explicitly considering the surfaces.

The design of simulation protocols to study ice growth from vapor is complex if the control of the vapor saturation at the ice surface and the existence of the QLL are considered. \citeauthor{mohandesi2018probing}\cite{mohandesi2018probing} introduced an MD simulation method to achieve steady-state ice growth from the vapor phase, using the TIP4P/2005 water model, which was accomplished by maintaining a high temperature near the ice surface to provide water molecules. This work emphasized that the behavior of the QLL and the growth rate are sensitive to the vapor flux. At low flux, growth is controlled by molecules reaching the ice interface, while at high flux, the deposition rate of water molecules can surpass the rate at which ice can grow at that temperature. This imbalance results in an inability to reach a steady state, and a liquid layer forms on the ice surface instead of solid ice, indicating that high vapor flux leads to a QLL that behaves more like liquid water. \citeauthor{libbrecht2014toward}\cite{libbrecht2014toward} used theoretical modeling and classical nucleation theory to investigate the mechanisms of ice crystal growth from vapor at the temperature near the melting point, focusing on the role of the QLL on the ice surfaces. They found that the QLL reduces nucleation barriers on the ice surface, facilitating the attachment of water molecules and crystal growth. For the basal face at low supersaturations, the step energy at the ice-QLL interface approximates that at the ice-liquid interface near the melting temperature, suggesting a direct correlation between ice crystallization mechanisms from vapor and liquid water. However, nucleation barriers do not significantly limit crystal growth on the prism surface or the basal surface at high supersaturations.
In this situation, the overall growth rate is predominantly controlled by the kinetic processes at the QLL-vapor interface. A study of ice growth at 260 K, using the TIP5P-E water model for ice Ih surfaces of different sizes and mimicking the QLL-vapor interfaces with slabs in contact with vacuum, confirmed that the basal surface has the lowest growth rate compared to the other investigated prismatic surfaces; differences were analyzed in terms of transient structures during the crystallization process \cite{doi:10.1063/1.4759113}. The coarse-grained mW model\cite{molinero2009water} was used to elucidate how supercooled water molecules are incorporated into the ice basal face in a slab, highlighting the transition from layer-wise to adhesive growth, implying roughness in a small temperature range close to the melting point \cite{mochizuki2023microscopic}. The mW model was also employed in simulations of slabs in the grand canonical ensemble to study condensation and evaporation\cite{pickering2018} and the development and surface structure of the QLL\cite{qiu2018}. Roughening of the QLL-vapor and the QLL-ice interfaces was also assessed\cite{benet2016premelting,baran2024possible}, and the phase features of the surfaces were related to the morphology habits with temperature\cite{llombart2020surface,berrens2022effect}. The roughness of the ice surface affects the adsorption of reactive species. It is most likely to be an important factor that influences the growth rate of the differently oriented crystal faces, thus controlling the shape of crystals formed from vapor.
Moreover, mesoscopic models were proposed\cite{kuroda1982,kuroda1990}, 
including a recent model for liquid-film mediated ice growth to identify distinct regimes at different levels of saturation\cite{sibley2021ice}. 

On the other hand, research based on MD on the growth of ice from melt has flourished, with selected representative efforts mentioned in the following. \citeauthor{NADA1996587} presented the first simulation study involving the discussion of ice Ih growth kinetics and mechanisms \cite{NADA1996587}. They proposed that ice grows following a {\it ``bilayer-by-bilayer''} mechanism on the basal surface, but according to a {\it ``collected molecule''} process on the primary prism surface, subsequently analyzing the translational and orientational order of the oxygen atoms to provide a detailed explanation of the two types of growth mechanisms\cite{doi:10.1021/jp963173c}.
\citeauthor{matsumoto2002molecular}\cite{matsumoto2002molecular} suggested that the key to ice growth relies on long-lived and sufficiently numerous hydrogen-bonded stable structures.
\citeauthor{doi:10.1080/00268970500243796}
\cite{doi:10.1080/00268970500243796} compared MD study of hexagonal ice growth from pure water and a brine solution in bulk systems, with the supercooling considered up to -18 K; they found that the ice growth is faster on the prismatic plane than on the basal plane. \citeauthor{NADA2005242}\cite{NADA2005242} also found that the growth rate on the second prismatic plane was more significant than that on the primary prismatic and basal planes, although they only explored one temperature, at 268 K. \citeauthor{doi:10.1080/00268970500243796}\cite{doi:10.1080/00268970500243796} and \citeauthor{NADA2005242}\cite{NADA2005242} both argued that the prismatic plane at the ice-liquid interface exhibits a geometrically rough state during ice growth, in contrast to the molecularly flat state of the basal plane. \citeauthor{10.1063/1.3451112} did extensive work developing the mW model \cite{molinero2009water} and assessed the reliability of applying this model to ice growth kinetics studies\cite{10.1063/1.3451112}. They argued that critical ice nuclei with less than ten water molecules develop faster than necessary for the liquid to relax, implying that supercooled liquid water cannot be adequately equilibrated; this has guided subsequent ice growth studies using this coarse-grained water model. \citeauthor{10.1063/1.3518984}\cite{10.1063/1.3518984} carried out simulations of ice growth from water at different interfaces of ice Ih and cubic (Ic) ice in bulk systems using the TIP4P, TIP4P-Ew and SPC/E water models, with the various water models revealing higher growth rates for both prismatic faces than for the basal face in the -65 to -40 K supercooling interval. \citeauthor{rozmanov2011temperature}\cite{rozmanov2011temperature} also examined the ice growth from water at ice-liquid interfaces in bulk systems, using the TIP4P-2005 water model, at temperatures ranging from -40 to +16 K with respect to the melting temperature. At around 12 K below the melting temperature, the growth rates approach a maximum value of 0.7 to 1.1 \AA/ns for systems with ``large'' ($2.7 \times 3.1$ nm$^2$) and ``small'' ($1.8 \times 1.6$ nm$^2$) interface size, respectively. The existence of a maximum in the growth rate as a function of temperature, predicted by experiments, has been confirmed by MD simulations in bulk in wider temperature ranges and with various water models\cite{weiss2011,Montero2019, montero2023kinetics}.

In summary, despite numerous studies, there is still no consensus on growth rates at different ice Ih surfaces both in experimental and theoretical/computational research, with a significant portion of studies focusing on narrow temperature ranges or bulk systems. Hence, further studies are needed to systematically determine how the ice growth rate varies over a wide range of supercooling temperatures in the presence of the QLL. As discussed previously, the existence of the QLL changes the bulk ice-liquid interface into the more complex ice-QLL-vapor double interface, and this structure mediates the absorption of water molecules from vapor and their incorporation into the crystal lattice.

The novelty of our work consists of investigating the kinetics of growth at surfaces including the QLL rather than at bulk liquid-ice interfaces in a wide temperature range for the surfaces relevant to growth from vapor. Moreover, we contribute to clarifying the evolution of the ice crystal morphologies between plate-like and column-like, which is linked to the temperature dependence of the growth rate.

\section{Methodology}
\subsection{Ice Slab Models}
Molecular dynamics simulations were performed using the rigid four-site TIP4P/Ice water model that was designed to study crystalline and amorphous ice phases and accurately reproduce the water phase diagram and the densities of several ice structures \cite{abascal2005general,abascal2007ice}; the ice Ih melting temperature was estimated at 272.2 K at 1 bar, which closely approximates experimental value \cite{abascal2005general}. However, even with the same TIP4P/Ice water model, the melting temperature calculated using different simulation techniques and details varies. More recent assessments estimate the TIP4P/Ice melting temperature at 269.8 K\cite{conde2017}, 268.78 K \cite{ji2024estimation}, 269.1 K\cite{bore2022phase}, and 270.3 K \cite{blazquez2022melting}. Hydrogens were initially in a disordered configuration determined by a Monte Carlo procedure to minimize the supercell dipole moment while obeying the Bernal-Fowler rule for hydrogen bonding \cite{grishina2004structure,kling2018structure}. The coarse-grained mW model \cite{molinero2009water}, which allows faster simulations, was also used for comparison.

\begin{figure}[htbp]
\centering
  \includegraphics[width=0.5\textwidth]{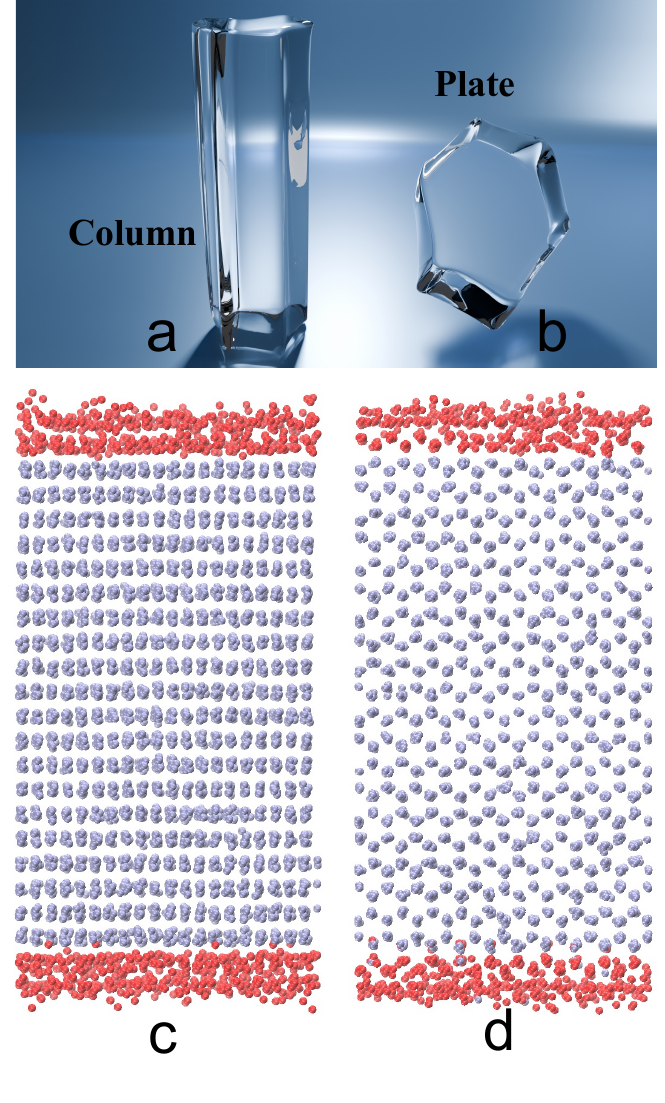}
  \caption{Column-like (a) and plate-like (b) morphologies for ice Ih crystals. If the growth along the basal surface direction is faster than along the prism1 surface direction, the crystal morphology would be column-like, while it would be plate-like if the reverse is true. The slab models used to study the basal (c) and primary prismatic (d) surfaces of ice Ih in an exemplary MD snapshot with TIP4P/Ice at 260 K. Only oxygen atoms are shown, with those liquid-like in red to highlight the QLL and the remaining in blue.
  }
  \label{model}
\end{figure}

Two ice Ih slabs composed of 5,760 water molecules were considered: one exposed the basal (0001) face to vacuum, while the other exposed the prism1 (10$\bar{1}$0) face. As periodic boundary conditions were applied in all three directions, a vacuum layer was added by elongating the dimension of the box by 40 \AA\ in the direction perpendicular to the exposed surface, producing two equivalent ice-vacuum interfaces. The size of the vacuum layer was chosen to be large enough so that the two ice-vacuum interfaces did not interact. Similarly, the length of the slab was chosen so that the two QLLs did not influence each other, especially at the higher temperatures and when the QLL was enlarged, resulting, before equilibration, in a slab of the total size of $45.18 \times 46.98 \times 130.00$ \AA$^3$ for the basal surface and of $45.17 \times 133.00 \times 44.36$ \AA$^3$ for the prism1 surface, respectively, as shown in Fig.~\ref{model} for MD snapshots at 260 K. The basal slab consisted of 48 layers, each composed of 120 molecules, arranged in pairs with an intra-bilayer spacing of 0.93 \AA\ and an inter-bilayer spacing of 3.74 \AA; the primary prism slab also consisted of 48 layers of 120 atoms each, arranged in pairs, with an intra-bilayer spacing of 1.30 \AA\ and an inter-bilayer spacing of 2.77 \AA. The supercells are oriented so that the basal surface is orthogonal to the Z direction, while the prism1 surface is orthogonal to the Y direction.
\subsection{Molecular Dynamics Simulations}
We started from the ice slabs in the QLL equilibrium state obtained by MD simulations as previously described\cite{shi2022investigating}. MD simulations were carried out for up to 160 ns, using the Large-scale Atomic/Molecular Massively Parallel Simulator (LAMMPS, Mar 03, 2020 version) package\cite{plimpton1995fast} in the NVT ensemble. For the TIP4P model, the cutoffs for the long-range Coulomb and Van der Waals interactions were set at 12 \AA, and a time step of 1 fs was used. For the mW model, a time step of 1 fs was also used for the crystallization simulations, while a 5 fs time step was used for the preparatory MD simulations to equilibrate the slabs and their QLLs. The smallest time step for mW was used because of the very fast crystallization rate with this force field so as not to lose details. 

The temperature was controlled by a chain of three Nos\'e-Hoover thermostats, with a relaxation time of 100 fs (equal to 100 time steps) at 240 K, 245 K, 250 K, 255 K, 260 K, 265 K, and 270 K for TIP4P/Ice and at 200 K, 210 K, 220 K, 230 K, 240 K, 245 K, 250 K, 255 K, 260 K, 265 K, and 270 K for mW. The reliability of the thermostat and its parameters were assessed by comparison with runs with a ten-fold larger time constant, the Langevin thermostat, and the canonical sampling through velocity rescaling thermostat\cite{bussi2007}. The comparison of the time evolution of potential energy, total energy, and temperature using different thermostats and different time constants is included in the Supplementary Material (SM), Fig.~S1-S5. We did not observe any artifact (e.g., as reported in the context of water crystallization in nanopores\cite{zhu2023}). Because the MD simulations were thermostatted, no heat dissipation was considered. This issue had been previously investigated by \citeauthor{Montero2019}\cite{Montero2019}; they calculated the growth rate at the secondary prism face in bulk with NVE MD simulations, which require large systems to equilibrate the temperature, obtaining the same results of previous thermostatted simulations (e.g. \cite{weiss2011}): this suggests that heat dissipation does not play a significant role in the dynamics of ice growth.

To imitate the conditions in which the growth kinetics are determined by the incorporation of water molecules in the ice lattice from the QLL, we artificially melted additional layers so that the size of one of the two QLLs in the slab was substantially larger (about 32 \AA) than at equilibrium and then observed it recrystallize during MD simulations. For the TIP4P model, the enlarged QLL was achieved by rapidly melting the additional bilayers for 2 ns at 300 K at one end of the ice slab while keeping the remaining atoms fixed at their original places. In practice for the TIP4/Ice simulations, the number of additional melted bilayers was 7 for 240~K - 255~K, 6 for 260~K - 265~K and 5 for 270~K.
After melting, the temperature of the QLL-ice region was returned to our target supercooled temperature. The atoms in the ice region were kept rigid, and this process was run for 1 ns. The QLL-ice interface was gently relaxed with all constraints removed by first performing a short MD with time steps ranging from 0.01 to 1 fs (100,000 steps with 0.01 fs, 100,000 with 0.1 fs, and 10,000 with 1 fs). NVT simulations were then carried out for up to 160 ns until the QLL returned to equilibrium upon recrystallization. To improve statistics, three repeat simulations were performed at each temperature. The MD replicas use different random number seeds to initialize the velocities in the QLL melting and QLL-ice interface relaxation stages. For 260 K and 265 K we simulated a fourth replica to ensure the results would not be significantly affected by statistics around critical crossovers.

The simulation protocol for the QLL crystallization simulations with the mW model was similar to that with TIP4P/Ice. However, we observed that the systems tended to recrystallize during equilibrium at the target temperature, so it was difficult for this coarse-grained water model to complete the relaxation protocol without starting the crystallization process \cite{molinero2009water}. Simulations of the QLL recrystallization were carried out for up to 3 ns for mW, given that the QLL crystallization happened substantially faster than with TIP4P/Ice. Ten mW replicas were simulated at each temperature.

We used the local Steinhardt order parameter $lq_3$ as implemented in PLUMED 2.7.3 \cite{tribello2014plumed, bonomi2009plumed} to distinguish between ice-like and liquid-like molecules with a cutoff for the first nearest neighbors of 3.5 \AA\ and a threshold value of -0.67. The $lq_3$ order parameter shows a favorable ratio between accuracy and computational efficiency with respect to other classification methods, which would, in any case, give qualitatively similar trends, as previously discussed\cite{shi2022investigating}.

The open-source software Blender (version 4.1.1)\cite{blender411} was used to draw exemplary ice crystal plate-like and column-like morphologies.

\section{Results and Discussion}
\subsection{Temperature dependence of the growth rate at the QLL-ice interface}
\begin{figure*}[htbp!]
\centering
\includegraphics[width=1.0\textwidth]{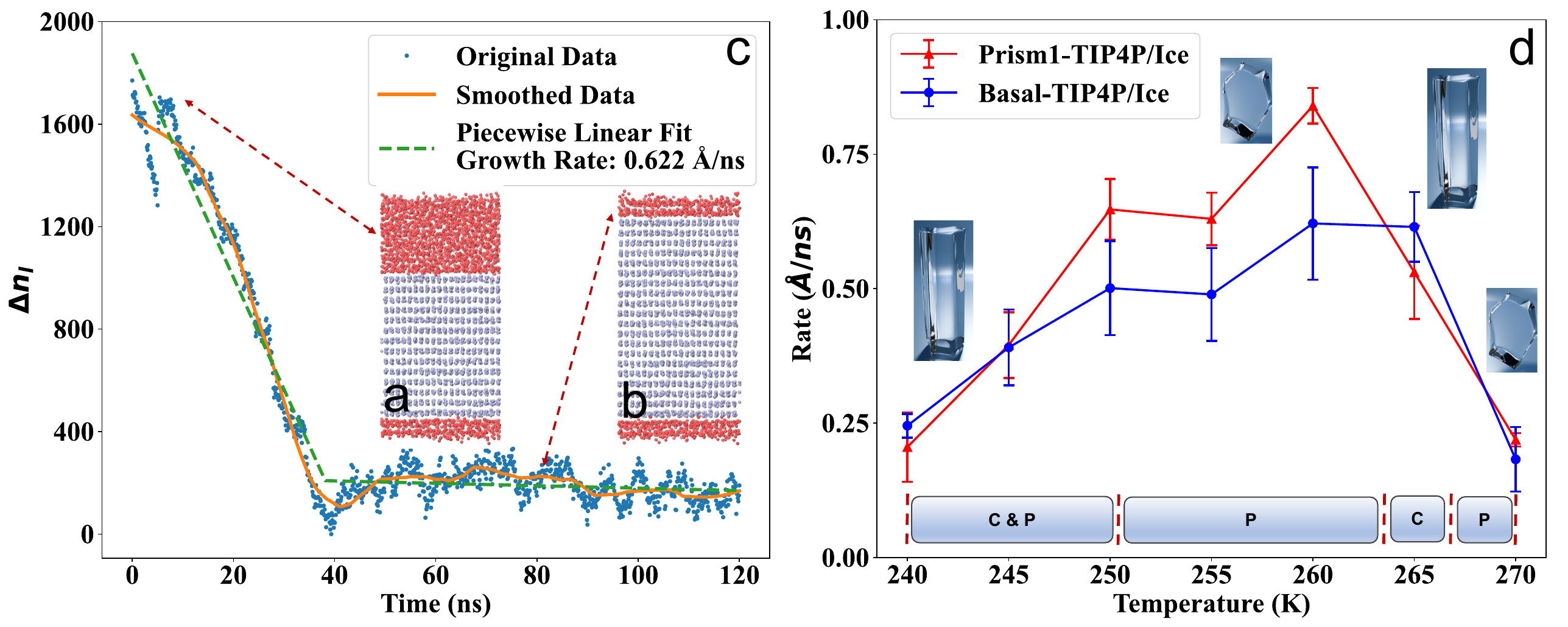}
\caption{The initial (a) and the final (b) structures in a 120 ns recrystallization simulation at 260 K for the basal face. The liquid-like molecules are colored red, and the ice-like molecules are blue. Only oxygen atoms are shown. Time-evolution (c) of the number of excess liquid-like molecules $\Delta n_l$ during the TIP4P/Ice crystallization simulations of the ice Ih basal surface at 260 K and illustration of the piecewise linear fitting technique used to obtain the ice growth rates. Growth rates (d) for the basal (blue) and prism1 (red) surface with TIP4P/Ice, averaged over replicas. The growth rates were determined without considering explicitly the saturation of water vapor on the QLLs, consistently with our model. Suggested morphologies compatible with our rates are shown in (d); the morphology images do not intend to represent correct dimensions. 
The vertical dashed lines in (d), delineating temperature intervals, are based on the Nakaya morphology diagram \cite{bender1962ukichiro} (labels C and P are column-like and plate-like, respectively), originally built by Ukichiro Nakaya 
and coworkers through experimental observations in the 1930s\cite{furukawa2007snow,libbrecht2005physics}; variations of these habits at low temperatures have been proposed\cite{bailey2009}.}
\label{envolution}
\end{figure*}

To estimate the ice growth rate for the different ice Ih surfaces, we monitored the time-evolution of the number of liquid-like molecules $n_l$ determined with the $lq_3$ order parameter and related it to the speed at which $n_l$ decreases in time. Although this is not a unique method, the $n_l$ evaluated with other appropriate classification methods would provide similar trends\cite{shi2022investigating,deepice}, and other suitable observables for rate calculations, like the potential energy or total energy\cite{zhu2023,weiss2011}, showed a time-evolution which strongly correlate with that of $n_l$. 

Fig.~\ref{envolution} c shows the time-evolution of the excess number of liquid-like molecules ($\Delta n_l$) on the ice Ih basal surface using the TIP4P/Ice model. $\Delta n_l$ is obtained from the simulations using $\Delta n_l(t)=n_l(t)-\min(n_l)$. The inner images a and b illustrate the recrystallization process at 260 K for the basal surface; similar data for the prism1 surface are shown in the SM, Fig.~S6. The initial liquid-like region (in red) on the top surface, which has been artificially melted, is considerably larger than the equilibrium QLL on the bottom surface (Fig.~\ref{envolution} a). The final configuration of the top surface at the end of the simulation (b) is similar to the bottom surface, confirming that crystallization has occurred and the melted QLL has returned to equilibrium. The decrease in $n_l$ is continuous, and its trend is approximately linear. 

The time-evolution of $\Delta n_l$ is also shown in the SM Fig.~S7 for both basal and prism1 surfaces, and both TIP4P/Ice (Fig.~S7a and S7b) and mW (Fig.~S7c and S7d) at each simulated temperature for one of the replicas. The mW crystallization process is remarkably rapid at a time scale between 1 and 2 ns, with respect to 160 ns for TIP4P/Ice. Crystallization occurs even during the relaxation period, consistently with findings by \citeauthor{molinero2009water}\cite{molinero2009water}, arguing that supercooled water using the mW model cannot be properly equilibrated due to the rapid crystallization process during the relaxation stage. In both water models, the time-evolution curves fluctuate considerably at the highest temperature of 270 K, which is very close to the melting point.

The TIP4P/Ice and mW growth rates were estimated using a piecewise linear fitting technique from the $\Delta n_l$ slope. The piecewise fitting procedure consists of three steps: the first step uses the Savitzky-Golay filter \cite{savitzky1964smoothing} (using polynomial fitting within each 200 data window) to smooth the raw data; the second step uses a dynamic programming method \cite{killick2012optimal} to find the breakpoint where the slope of the curve changes significantly; the third step divides the fitting interval according to the breakpoint. The first decreasing trend fitting curve (the fitting curve before the breakpoint) is used to obtain the ice growth rate. The slope from the fitting is then converted into a volumetric growth rate and eventually into a thickness growth rate. The thickness $d$ in \AA\ was obtained as 
$$d = \frac{\Delta{n_l} M}{\rho N_{\mathrm{AV}} S 10^{-24}}{\rm ,} $$ 
as previously used to evaluate the QLL thickness from the number of liquid-like molecules\cite{shi2022investigating}; here $M$ = 18.01574 g mol$^{-1}$ is the molecular weight of water, $N_{\mathrm{AV}}$ is the Avogadro number, $\rho$ is the density of water in units of g cm$^{-3}$, and $S$ is the size of the ice surface exposed to vacuum. This fitting procedure significantly reduced the errors caused by an arbitrary artificial selection of the start and end points for the fitting; it is an upgrade on the linear fitting employed, for example, in Ref.\cite{rozmanov2011temperature} and Ref.\cite{shi2024}. Only data for the growth process on the basal surface at 260 K are shown in Fig.~\ref{envolution} c as an example; further data and fitting at different temperatures and surfaces are shown in the SM, Fig.~S8-S14. The temperatures at the extremes (240 K and 270 K) are those that behave less regularly due to the unbalance between the competing effects in the growth process. 
\begin{figure}[htbp!]
\centering
\includegraphics[width=0.48\textwidth]{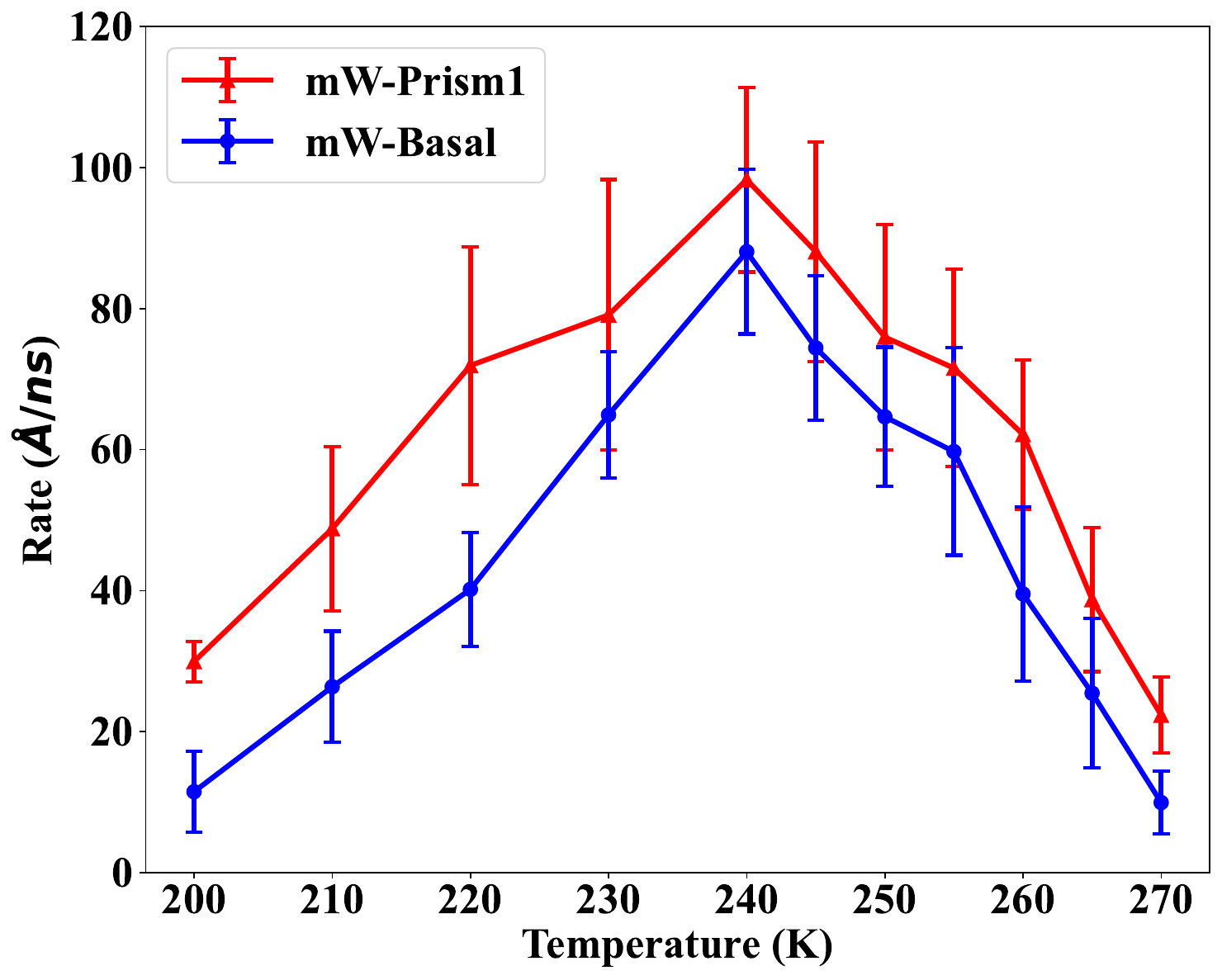}
\caption{Growth rates with temperature from 200 K to 270 K for the basal (blue) and prism1 (red) surfaces of the mW model, averaged over replicas.}
\label{rate}
\end{figure}

The TIP4P/Ice ice growth rates of the basal and prism1 surfaces as a function of temperature are shown in Fig.~\ref{envolution} d. The mW growth rates of these two surfaces are shown in Fig.~\ref{rate}. The rates were averaged over three replicas for TIP4P/Ice (four for 260 K and 265 K) and ten replicas for mW. The error bars indicate the standard deviations of the replicas. The crystallization rate shows a gradual increase and then a gradual decrease with temperature. It reaches a maximum of 260 K (TIP4P/Ice) and 240 K (mW) for both basal and prism surfaces. This is due to the interplay between the interface chemical potential and diffusion, which have opposite trends with temperature \cite{Montero2019} as explained in more detail below. We observe differences between the temperature dependence of the growth rate of mW with respect to that of the TIP4P/Ice model.

For the TIP4P/Ice simulations, low growth rates on different ice surfaces are found at both low temperatures of 240 K and high temperatures of 270 K (Fig.~\ref{envolution}). At the highest temperature of 270 K, as shown in the SM  Fig.~S14, the excess number of liquid-like molecules in the QLL shows fluctuations consistent with the proximity to the melting temperature. A high level of supercooling leads to an easier transition from water to ice. However, at low temperatures, the thermal movement of molecules is decreased, so the clustering, collision, and ordering of molecules needed for ice growth are weakened, leading to a reduction in the growth rate. On the other side of the range, at low supercooling and in proximity to the melting point (270 K), the substantial thermal movement of the molecules causes the newly formed regularly arranged ice structure to disintegrate. This slows down the ice growth rate. A similar scenario also applies to the case of the mW model at low and high supercooling, as shown in Fig.~\ref{rate}, although mW peaks at 240 K, which is different from TIP4P/Ice. In the mW model, the absence of explicit hydrogen atoms leads to lower diffusion energy barriers and much faster dynamics \cite{molinero2009water}, resulting in significantly larger rates of QLL crystallization compared to TIP4P/Ice. Overall, the growth rate of the prism1 surface is greater than that of the basal surface over a wide range of temperatures (always in the case of mW). To clarify the reasons for the faster crystallization rate of the prism compared to the basal at some temperatures \cite{10.1063/1.3518984,NADA2005242}, we explored the ice growth mechanisms on different ice surfaces below.
\begin{figure}[htbp!]
\centering
\includegraphics[width=0.48\textwidth]{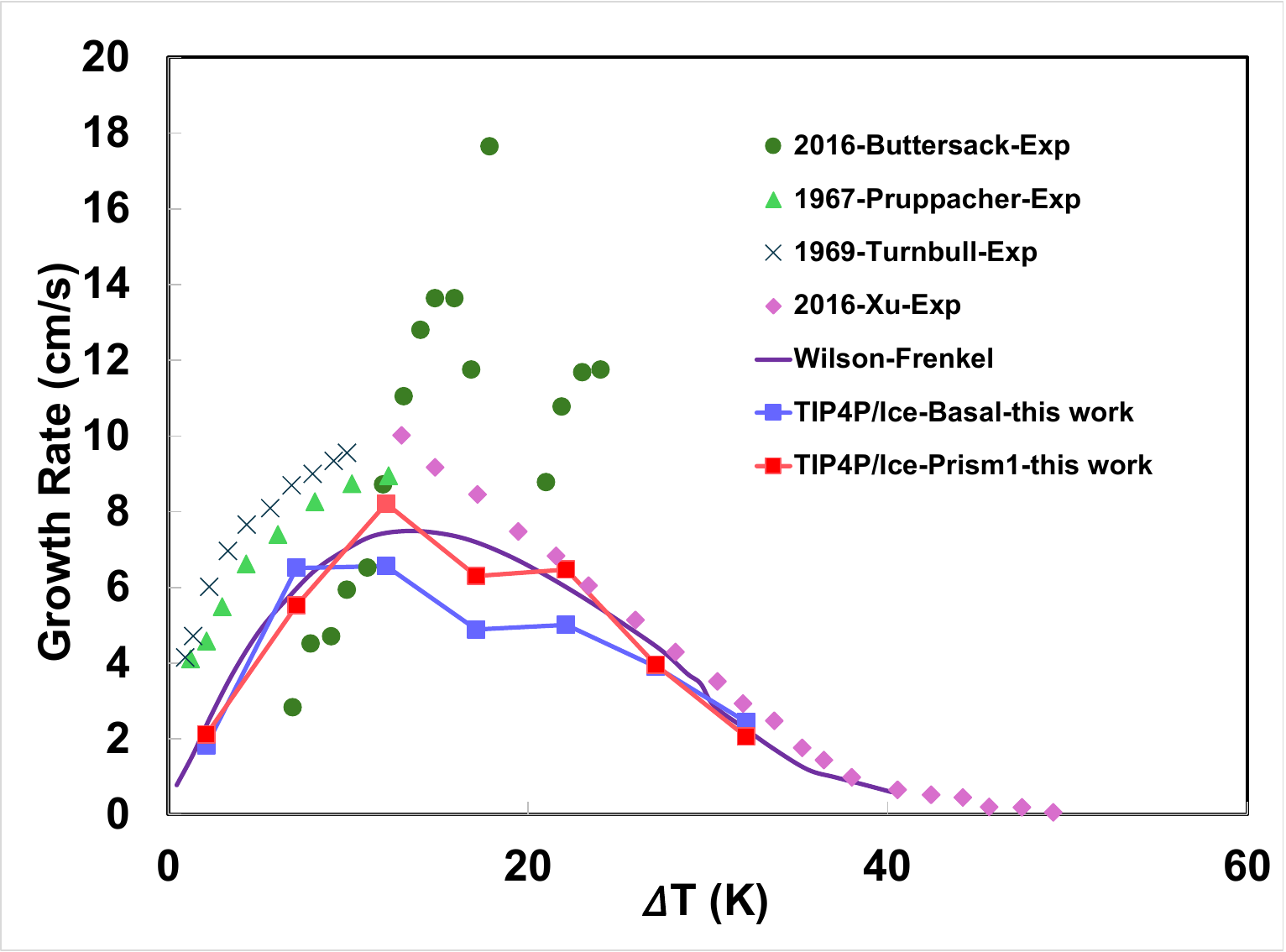} 
\caption{Comparison of ice Ih growth rates as a function of supercooling of our simulation data for the basal (blue line) and prism1 (red line) surfaces, the WF model with parameters for the TIP4P/Ice force field (purple line) and experimental results. The experimental data are from \citeauthor{buttersack2016critical} \cite{buttersack2016critical} (dark green dots), \citeauthor{pruppacher1967interpretation}\cite{pruppacher1967interpretation} (green triangles), \citeauthor{turnbull1969under}\cite{turnbull1969under} (grey crosses), and \citeauthor{xu2016growth}\cite{xu2016growth} (pink diamonds).}
\label{comparison}
\end{figure}

The growth rates from our TIP4P/Ice simulations range from ~0.2 to 0.9 \AA/ns for both the basal and prism1 surfaces at different temperatures. At 260 K, they reach the highest at 0.621$\pm$0.104 and 0.840$\pm$0.033 \AA/ns for the basal and prism1 surfaces, respectively. The existence of a maximum growth rate below the melting point of ice is predicted by experiments and has been previously confirmed in several bulk simulations, which show similarities with our slab trends and values of rates. For example, \citeauthor{rozmanov2011temperature}\cite{rozmanov2011temperature} evaluated the effects of different system sizes on the growth rates of the basal plane in the bulk with the TIP4P-2005 model, finding values ranging from 0.7 to 0.9 to 1.1 \AA/ns at 10 K below the melting point for their large, medium and small systems. \citeauthor{Montero2019} \cite{Montero2019} studied the ice growth rate for the secondary prismatic plane in the bulk with the TIP4P/Ice model. They found growth rates of about 1 \AA/ns at -14 K and 0.5 \AA/ns at -25 K, with values in line with our simulations focusing on the basal and prism1 surfaces. These rates were obtained without the use of thermostats with results consistent with thermostatted simulations (including \citeauthor{weiss2011}\cite{weiss2011}).
The maximum ice growth rate at a certain degree of supercooling also occurs in the mW model, with our simulations showing that the highest mW growth rate occurs at 240 K. A similar behavior was also found for the mW growth rates of the prism2 face by \citeauthor{espinosa2016time}\cite{espinosa2016time}, with the highest value at about~234.6 K (corrected rate values\cite{montero2023kinetics}) and by \citeauthor{guo2022}\cite{guo2022} (who also calculated the rates for TIP4P-2005). A recent comparison of rates in the bulk for various models was reported by \citeauthor{montero2023kinetics}\cite{montero2023kinetics}, with the rates for a machine learning potential sitting in between mW and TIP4P results. In our slab simulations of the basal and prism1 surfaces, we found mW growth rates two orders of magnitude larger than those of TIP4P/Ice.

The Wilson-Frenkel (WF) model \cite{wilson1900xx,frenkel1955kinetic} includes the effects of the competitive mechanisms leading to the growth rate trend with respect to temperature; it considers diffusive order kinetics as governed by the diffusion coefficient, which is assumed proportional to the addition rate. The predicted growth rate $R(T)$ is expressed as
\begin{equation}
R(T)=(D(T) / \alpha)\left[1-\exp \left(-\Delta G_{lx}(T) / k_b T\right)\right]
\end{equation}
where $D(T)$ is the diffusion coefficient, $\Delta G_{lx}(T)$ the chemical potential difference between the solid and the liquid phases, $\alpha$ a molecular dimension (the diameter of a water molecule) and $k_B$ the Boltzmann constant. We plotted in Fig.~\ref{comparison}, the results for the WF model with parameters, as $D(T)$ and $\Delta G_{lx}(T)$, from TIP4P/Ice simulations in bulk system and $\alpha = 3$ \AA \cite{espinosa2016time,Montero2019}. $\Delta G_{lx}(T)$ can be assumed to have linear behavior from a value calculated at a reference temperature going to zero at the melting point \cite{espinosa2016time,Montero2019}.

Our results for the slabs, shown for comparison in Fig.~\ref{comparison}, are qualitatively similar in the trend but also capture differences for the different surfaces. Here, the growth rate unit is in cm$/$s=10$^{-1}$\AA$/$ns. Our simulation results are consistent with experiments on nanofilms\cite{xu2016growth} at large supercooling, particularly for the growth rate of our prism1 surface. However, the simulated results do not agree with those of \citeauthor{buttersack2016critical} \cite{buttersack2016critical}, who showed significantly higher growth rates 
in experiment work of droplet nucleation in vapor. Our simulations showed the same increasing trend at moderate supercooling as \citeauthor{turnbull1969under}\cite{turnbull1969under} and \citeauthor{pruppacher1967interpretation}\cite{pruppacher1967interpretation} experimental results, except that their experimentally measured growth rate values were slightly larger than ours. We did not reproduce the linear behavior close to the melting point experimentally observed by \citeauthor{lamb1972linear} for ice crystal growth from vapor. However, they also recorded maxima as in our and other simulations and experiments. We reiterate that our rates are defined in the approximation where the thermostat makes the system isothermal, although results in bulk do not seem to be strongly affected\cite{espinosa2016time, weiss2011,Montero2019}. In experiments, the latent heat, released at the interface as crystallization proceeds, must diffuse away from the interface to allow further crystallization, as thermal energy build-up can hinder further crystallization 
\cite{sanchez2017experimental}; our simulations do not account for this and, with respect to the bulk, the QLL maintains a diffusive and disordered layer at equilibrium.

Although with caution due to the calculation uncertainties, the TIP4P/Ice growth rate can be qualitatively correlated with the changes in the morphology of ice crystals, as illustrated in Fig.~\ref{envolution} d. When the ice growth rate on the basal plane is larger than the rate on the prism1 plane, ice crystals will grow with a columnar shape, while when the prism1 rate is larger than the basal one, they will grow like platelets. In the TIP4P/Ice simulations, three crossovers in the average rate can be observed. The first crossover occurs close to 245 K, where the rates of basal and prism1 are very similar, with a potential coexistence of plate-like and column-like crystals. Then, as temperature increases, the prism1 rate is larger than the basal one, favoring plate-like shapes. Two switches to column-like and plate-like occur in a small temperature range around 265 K. Although caution has to be considered given the error bars and limitation of the water model and the simulation protocol, these trends showing an alternation of favorable shapes are qualitatively consistent with the morphology diagram based on experimental evidence \cite{libbrecht2005physics,libbrecht2017physical,furukawa2007snow,bailey2009}.
From the time-evolution of the replicas (reported in the SM), the basal rates seem to have more variations than the prism1 at 260 K. Vice versa at 265 K, showing some change of behavior at this point, with the average rates crossing, or becoming comparable if accounting for the uncertainties, but in any case showing a change from the largest prism1 rates between 250 and 260 K.
Crossovers in the thickness of the QLL and the average values of order parameters were also observed in our previous equilibrium MD simulations \cite{shi2022investigating}, particularly at 265 K and 245-250 K. However, it is not straightforward to correlate them with morphology habits.

In the case of mW, the prism1 surface grows faster than the basal one at all temperatures, suggesting the growth of platelet-shaped ice crystals at all temperatures. Our mW MD simulations also confirmed that the QLL thickness on the basal and prism1 ice Ih surfaces increases with temperature, with no crossovers between each other. With mW MD simulations, \citeauthor{mochizuki2023microscopic}\cite{mochizuki2023microscopic} demonstrated a shift in the growth mode of the basal face within less than 5 K from the melting temperature (our simulations do not have this temperature resolution), from bilayer-by-bilayer to adhesion growth mode, highlighting the critical role of interfacial roughness. 

Recent studies \cite{llombart2020surface,berrens2022effect} have looked at the question of morphology habits from a different perspective, not directly calculating rates but analyzing changes in properties of the surfaces with temperature and correlating them with changes in crystal shape.
\citeauthor{llombart2020surface}\cite{llombart2020surface} quantified the roughening transition of ice surfaces at different temperatures through equilibrium TIP4P/Ice MD simulations for systems with large surfaces, revealing how temperature-driven interfacial phase transitions affect ice growth rates. They identified the role of phase transitions at the surface, specifically from an ordered flat (OF) to a disordered flat (DOF) phase as temperature changes. As the temperature increases, the phase transformation from DOF-like to a high-temperature reconstructed flat (HT-RF) state exhibits anomalous step-free energy increases. This sequence of transitions explains the crossover observed in crystal growth rates of the basal and prism facets of ice, correlating with the formation of plate-like and columnar crystal habits seen in the atmosphere. \citeauthor{berrens2022effect}\cite{berrens2022effect} investigated (with TIP4P/2005) the effects of $\mathrm{Na}^{+} / \mathrm{Cl}^{-1}$ on the ice surface properties, and found that high concentrations of ions significantly increased the QLL thickness and roughness, thereby affecting the kinetics of crystal growth.
\begin{figure*}[htbp!]
\centering
  \includegraphics[width=0.85\textwidth]{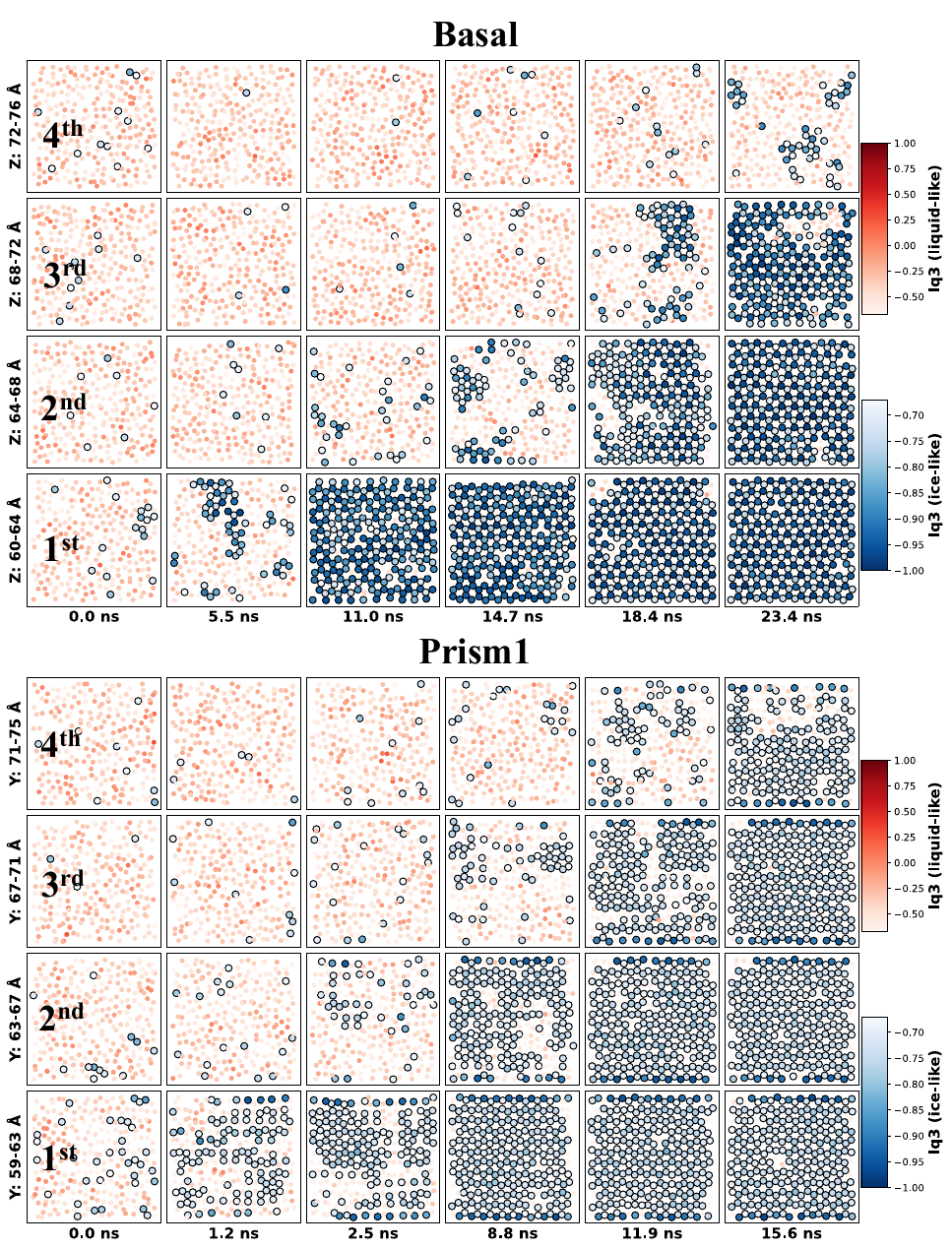}  
  \caption{Time-evolution of the formation of the four ice bilayers above the interface where the melted region initially met the crystalline region during the TIP4P/Ice crystallization process at 260 K. The darker blue indicates that the molecules are in a more ordered state, i.e., ice-like molecules; the more reddish molecules correspond to a more disordered state, i.e., liquid-like molecules. The oxygen atoms are colored according to the values of the order parameter $lq_3$.}
  \label{tip4p_kinetics}
\end{figure*}

The habit diagram from laboratory studies\cite{bailey2009} includes plates (in a range from the melting temperature from 0 to -4 K), to column (-4 K to -8 K), to plates (-8 to -22 K), agreeing well with typical habit diagrams in atmospheric science\cite{furukawa2007snow,libbrecht2005physics}. At lower temperatures, complex polycrystals rather than single crystals were found, 
suggesting plate-like shapes from -20 to -40 K and columnar from -40 K to -70 K\cite{bailey2009}. As MD-based simulations only consider surfaces, implying single crystals, only the first three cross-overs can be meaningfully considered. The crossovers of our average growth rates with respect to 270 K can only be identified within 5 K temperature intervals, given that the calculations were carried out every 5 K, plus the calculation uncertainties; they are around -5 to 0 K, -10 to -5 K, and –30 to -20 K, which suggest correlation with ice morphology changes between plate-like, column-like and plate-like again, with good qualitative agreement with experiments\cite{bailey2009}. \citeauthor{llombart2020surface}'\cite{llombart2020surface} results show a first crossover at -6 K from the melting temperature, with a plate-like morphology at higher temperatures, and a second transition (columns to plates) at -26 K\cite{llombart2020surface}. \citeauthor{berrens2022effect}'\cite{berrens2022effect} simulation results also identified the temperatures related to morphology changes based on the sequence of roughening transitions, predicting for pristine ice, as temperature decreases from the melting point, a transition from columns to plates at -15 K, from plates to plates and/or columns at -25 K, and to columns again between -39 K and -49 K. In summary, there is some qualitative agreement in determining temperature crossovers and relating them to crystal habits within different computational approaches. However, the crossover temperature has some variations, which can be expected due to the different methodologies, water models, simulation protocols, statistical uncertainties, and approximations. As mentioned in Ref.\cite{llombart2020surface}, 
precise location of the crossovers could depend on different factors and the probed effects. Nevertheless, all these studies contribute to elucidating, at the molecular level, the growth and resulting shapes of ice crystals.  
\begin{figure*}[htbp!]
\centering
\includegraphics[width=1.0\textwidth]{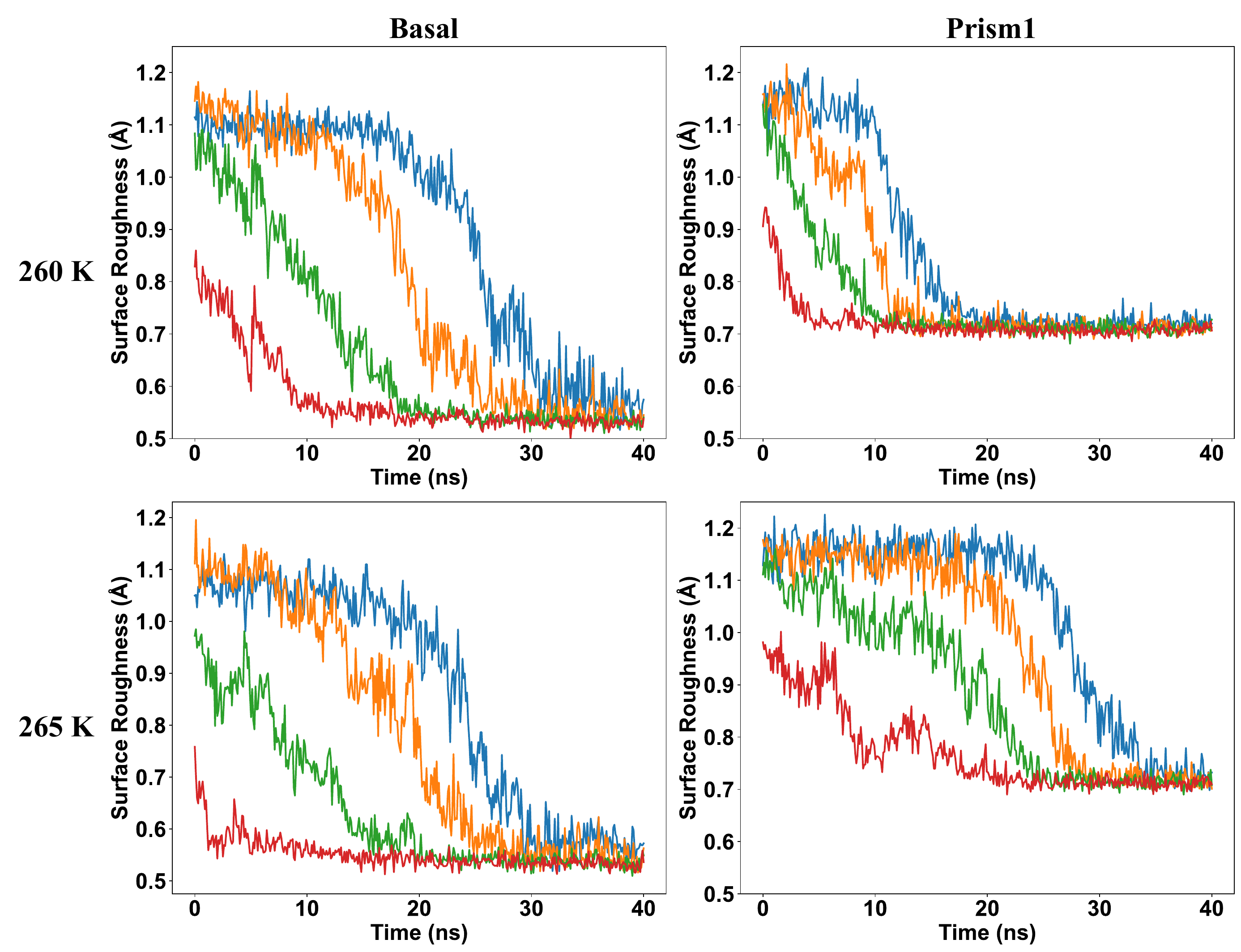} 
\caption{Time-evolution of the roughness of the four bilayers above the interface where the melted region initially meets the crystalline region for the basal (left panel) and prism1 (right panel) faces during the TIP4P/Ice crystallization at 260 K and 265 K.}
\label{roughness}
\end{figure*}

\begin{figure*}[htbp!]
\centering
  \includegraphics[width=1.0\textwidth]{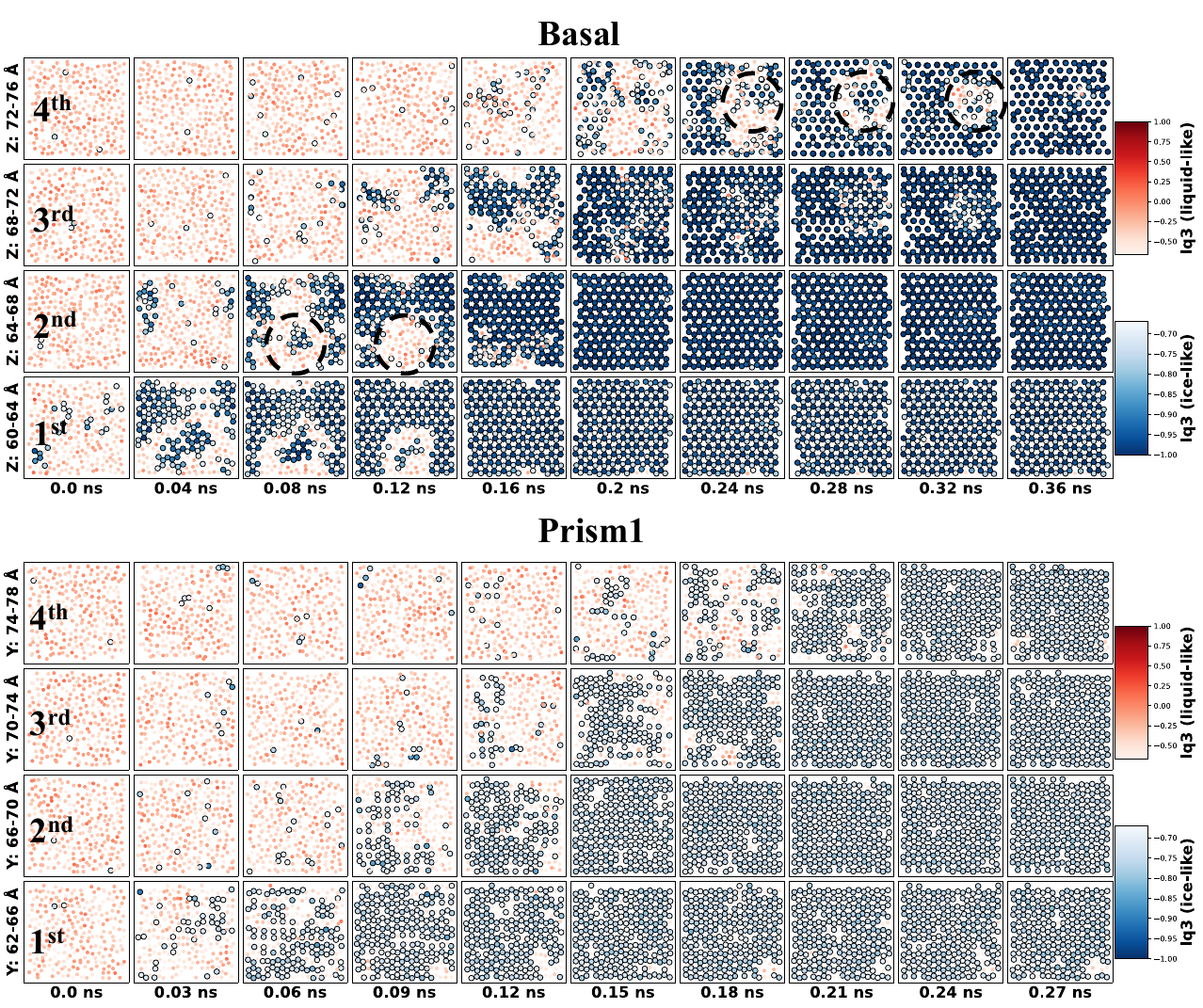}  
  \caption{Time-evolution of the formation of the four ice bilayers above the interface where the melted region initially met the crystalline region for the basal (top panel) and prism1 (bottom panel) faces during the mW crystallization process at 240 K. The darker blue indicates that the molecules are in a more ordered state, i.e., ice-like molecules; the more reddish molecules correspond to a more disordered state, i.e., liquid-like molecules. The oxygen atoms are colored according to the values of the order parameter $lq_3$.
  }
  \label{mw_kinetics}
\end{figure*}
\subsection{Growth mechanisms of TIP4P/Ice ice Ih at the basal and prism1 QLL-ice interface}
The formation of the four bilayers above the interface where the melted region initially met the crystalline region was investigated by monitoring the water molecules in that region. 
 In terms of the supercell coordinates, bilayers 1 to 4 correspond to 60-64 \AA, 64-68 \AA, 68-72 \AA\ and 72-76 \AA, respectively, in the Z direction orthogonal to the basal plane for the basal surface model, and to 59-63 \AA, 63-67 \AA, 67-71 \AA\ and 71-75 \AA, respectively, in the Y direction orthogonal to the prism1 plane for the prism1 surface model. The time-evolution of the formation of these four ice bilayers at the basal (top panel) and prism1 (bottom panel) faces during the TIP4P/Ice crystallization process at 260 K is shown in Fig.~\ref{tip4p_kinetics}. We found that the crystallization process on the basal plane follows a ``bilayer-by-bilayer'' mechanism\cite{NADA1996587}. The bilayer-by-bilayer pattern for ice crystal growth involves sequentially adding water molecules to the existing ice structure. This process occurs primarily through a mechanism in which molecules adhere to specific sites at the ice interface, forming a new layer that spreads laterally across the crystal. Growth proceeds uniformly as each layer completes before a new one starts. This can be seen in the first bilayer at 11 ns, the second bilayer at 18.4 ns, and the third bilayer at 23.4 ns. The first bilayer is ordered as the simulation progresses to 11 ns, while most molecules in the second bilayer remain disordered. As the simulation continues to 18.4 ns, the second bilayer completes its crystallization and becomes ordered. Although the third bilayer of water molecules is partially ordered with a few ice-like molecules, most of the area remains liquid-like. When the simulation reaches 23.4 ns, the third bilayer completes crystallization and becomes ordered, but the fourth bilayer only has a few ice-like molecules. These behaviors reflect the bilayer-by-bilayer growth mechanism on the basal surface. 

On the other hand, the prism1 plane appears to follow a ``collected molecule'' \cite{NADA1996587} mechanism, which refers to a process where molecules adhere to an interface in a stepwise manner, in agreement with observations by \citeauthor{NADA1996587}\cite{NADA1996587}. The structures at 8.8 ns, 11.9 ns, and 15.6 ns for the prism1 plane in the bottom panel in Fig.~\ref{tip4p_kinetics} reflect the ``collected molecule'' growth mechanism. At 8.8 ns, the first bilayer completely crystallizes and becomes ordered. At this point, a large part of the second layer also becomes ordered, and small amounts of ice-like molecules appear in the third and fourth bilayers. When the simulation progresses to 11.9 ns, the second bilayer is entirely crystallized, the third bilayer is essentially crystallized, and some ice-like molecules even appear in the fourth bilayer. When the simulation reaches 15.6 ns, the third bilayer has completed the crystallization process, and the fourth bilayer is also mostly crystallized. Through the above analysis and comparison with the growth mechanism of the basal plane, we can conclude that the ``collected molecule'' mechanism is a sort of three-dimensional growth mode, leading to a more localized and less uniform growth. Ice-like molecules are collected at the interface and arranged into a structured bilayer. This three-dimensional growth mechanism leads to the formation of ordered ice-like local structures in the space ``vertically'' above the currently crystallizing bilayer, which also significantly accelerates the growth rate of the prism1 face.


\subsection{Bilayer roughness of TIP4P/Ice ice Ih at the basal and prism1 QLL-ice interface}
As previously discussed, the degree of order/disorder of the basal and prism1 ice Ih surfaces has been related to the crystal morphology propensity through analysis of the roughness, which has been extensively investigated considering both the QLL-vapor and QLL-ice interfaces\cite{llombart2020surface,berrens2022effect,benet2016premelting,baran2024possible}. Here, we concentrated our analysis on the time-evolution of the roughness of the four bilayers described above for the basal and prism1 orientations during the process of growth, using the root mean square roughness \cite{khayet2005characterization}
\begin{equation}
    R(t) = \sqrt{\frac{1}{N(t)} \sum_{i=1}^{N} (z_i(t) - \bar{z}(t))^2}
\end{equation}
where $z_i$ is the ``height'' of oxygen atoms within the bilayer region in the direction orthogonal to the surface, $\bar{z}$ is the average height, and $N$ is the number of oxygen atoms within the bilayer region.

Fig.~\ref{roughness} shows the time-evolution of the roughness of the four selected bilayers in the basal and prism1 surfaces at 260 and 265 K at specific simulation times for representative replicas. The basal and prism1 planes initially exhibit a large roughness in the four bilayers because the first bilayer is still in the preparatory stage of crystallization, and the other bilayers are located further away in the upper part of the enlarged QLL, also exhibiting a large roughness. For the basal plane at 260 K, the first bilayer undergoes crystallization and flattens as the simulation continues, so its roughness decreases. In contrast, the roughness of the second bilayer decreases to a certain degree as the water molecules in this layer are required to form this neighboring ice layer. However, the third and fourth bilayers continue to show greater roughness. When the simulation reaches 18.4 ns, the roughness of the second bilayer decreases considerably to complete the crystallization process. Still, the roughness of the fourth bilayer remains unchanged, while the third bilayer shows a slight decrease in roughness. When the simulation reaches 20 ns, more parts of the region of the third bilayer has become ordered, i.e., the roughness decreases further. Still, the roughness of the fourth bilayer remains unchanged. The roughness of the basal plane further confirms that the growth of the basal plane tends to be more of a bilayer-by-bilayer growth pattern and does not perturb the roughness of the other layers significantly. The roughness of the four bilayers on the surface of prism1 in Fig.~\ref{roughness} right panel, on the other hand, shows a somehow different trend with time. At the start of the simulation, bilayers show a relatively large roughness because of the disordered local liquid-like environment. As the simulation proceeds, we can see that nearly all the bilayers show a simultaneous decreasing trend in roughness. This reflects that the growth process on the surface of prism1 is that all the molecules in the QLL region are actively involved in the crystallization process for each bilayer,
 which has a three-dimensional character.

The basal plane shows a mixed growth pattern combining ``bilayer-by-bilayer'' and ``collected molecule'' characteristics at 265 K compared to 260 K. The upper bilayers showed the emergence of locally ordered ice-like environments when the lower bilayers crystallized, which is why the second and third bilayers exhibited a significant decrease in roughness from 5 ns to 16 ns, which is consistent with the results in SM, Fig.~S15. However, the roughness of the third bilayer at 260 K remained high over a similar time range. On the other hand, the prism1 surface exhibited likewise a bilayer-by-bilayer growth mechanism that usually occurs for the basal face. After the roughness of the first bilayer significantly decreased, the third and fourth bilayers remained essentially the same with higher roughness, except for the second neighboring bilayer, which showed a certain degree of roughness reduction. This may explain why the growth rate of the basal plane was larger than that of the prism1 plane at 265 K. As the temperature varied, the growth rate of different ice surfaces changed, and this was reflected by changes in the growth mechanisms, which ultimately would lead to changes in the ice morphologies.

\subsection{Growth mechanisms of mW ice Ih at the basal and prism1 QLL-ice interface}
To explain the reason for the fast growth rate of mW, we analyzed the MD trajectories at specific simulation times for the two surfaces. We observed that the growth mechanism in both planes showed features of a three-dimensional growth pattern of ``collected molecules'', which explains to some extent why their growth rate is two orders of magnitude that of the TIP4P/Ice model. For the basal plane, the molecular regularization developed around some regions, highlighted by black circles as shown in Fig.~\ref{mw_kinetics}, showing molecular-ordered and disordered transformations at different times. 
In comparison, the bilayer on the prism1 surface shows homogeneous growth. That is to say that during the growth of each bilayer, defects appear uniformly everywhere in the plane instead of appearing first in some region enclosing a liquid-like region and then gradually expanding towards the enclosure to complete the crystallization process, as in the case of the basal plane. Thus, the mW prism1 plane still exhibits a faster growth process.

\section{Conclusions}
In this work, we systematically investigated crystallization at the ice-QLL interface for the Ih basal and prism1 surfaces with MD simulations using the TIP4P/Ice and mW force fields in a wide range of temperatures. To achieve this, we assumed that at low supersaturation, the rate-limiting mechanisms of growth are the integration of molecules from the QLL to the crystalline lattice and introduced an enlarged out-of-equilibrium QLL, which relaxes to equilibrium only through the incorporation of liquid-like molecules into the crystal. We acknowledge that other interpretations of the relevant supersaturation regime for growth from an out-of-equilibrium enlarged QLL are also possible\cite{sibley2021ice}.

The crystallization rates obtained with both water models increased with temperature, reaching a maximum; they then followed a downward trend toward the melting temperature, showing similarities with the crystallization rates from the melt as simulated in bulk systems rather than slabs. The TIP4P/Ice and the mW growth rates peak at 260 K and 240 K, respectively, with mW rates much faster than TIP4P/Ice ones. The experimental evolution of ice morphologies with temperature at low vapor pressure between column-like and plate-like can be qualitatively correlated with the ice growth rates in the case of TIP4P, although with caution given statistical uncertainties, but not for mW. Order parameters and QLL thickness also show variations with temperature with temperature that correlates with the change in ice morphology\cite{shi2022investigating}, which has also be linked with surface roughness transitions\cite{llombart2020surface,berrens2022effect}. 

The anisotropic ice growth mechanisms for the two QLL-ice interfaces of ice Ih were discussed, showing that for the TIP4P/Ice model at 260 K, the basal surface follows a ``bilayer-by-bilayer'' pattern, while the prism1 obeys a ``collected molecule'' pattern. For the basal surface, 
the roughness of the bilayer undergoing crystallization indicated a sequential crystallization process. For the prism1 surface, however, all four bilayers showed a simultaneous and consistent decrease in roughness over time. This suggested a three-dimensional growth mechanism where all layers were actively involved in the crystallization process. It is worth noting that the basal plane demonstrates a combination of these two distinct growth modes at 265 K, while the prism1 plane adopts the ``bilayer-by-bilayer'' pattern usually shown by the basal face.

For the mW model, the growth mechanism for both the basal and prism1 surfaces followed a three-dimensional ``collected molecule'' pattern, accounting for the significantly faster growth rate than for the TIP4P/Ice model. 
Although our protocol has limitations, there is a qualitative agreement with experimental prediction of morphology habits with temperature\cite{furukawa2007snow,libbrecht2005physics,libbrecht2017physical,bailey2009}, as well as MD simulations attributing crystal habits to changes in roughness and surface phase behaviour\cite{llombart2020surface,berrens2022effect}. Hence, it provides complementary information that contributes to understanding ice crystal growth.

\section*{SUPPLEMENTARY MATERIAL}
The Supplementary Material includes: 
the comparison of results obtained with the Nos\'e-Hoover thermostat with different time constants (Fig.~S1); the comparison of results obtained with the Nos\'e-Hoover and Langevin thermostats (Figs.~S2 and S3); the comparison of results obtained with the Nos\'e-Hoover and CSVR thermostats (Figs.~S4 and S5); the initial and the final structures in a recrystallization simulation at 260 K for the basal and prism1 surfaces (Fig.~S6); the time-evolution of the excess number of liquid-like molecules $\Delta n_l$ during the TIP4P/Ice and mW crystallization simulations for one replica at each temperature, for the basal and prism1 surfaces (Fig.~S7); the piecewise linear fitting of the time-evolution of $\Delta n_l$ during the TIP4P/Ice crystallization simulations for the replicas of the ice Ih basal and prism1 surfaces from 240 K to 270 K (Fig.~S8-S14); the time-evolution of the formation of the four ice bilayers above the interface where the melted area initially met the crystalline region for the basal and prism1 faces during the TIP4P/Ice crystallization process at 265 K (Fig.~S15). 

\begin{acknowledgments}
We are grateful for computational support from the UK high-performance computing services ARCHER2, for which access was obtained via the UKCP consortium and funded by EPSRC grants EP/P022472/1 and EP/X035891/1, and the UK Materials and Molecular Modelling Hub for computational resources, which is partially funded by EPSRC (EP/T022213/1). We thank Alejandro Santana Bonilla (King's College London) for technical support. JS was supported by a King's China Scholarship Council PhD studentship.
\end{acknowledgments}

\section*{AUTHOR DECLARATIONS}

\noindent{\bf Conflicts of interest}

There are no conflicts of interest to declare.

\noindent{\bf Author Contributions}
JS, MF, MS, and CM participated in the research design. JS and MF performed the simulations. JS analyzed the data. JS and CM wrote the manuscript. All authors contributed to the discussion of the results and the revision of the manuscript and approved the submitted version.

\section*{DATA AVAILABILITY STATEMENT}
The data supporting this article is openly available from the King’s College London research data repository, KORDS, at https://doi.org/10.18742/27080410.
\appendix

\nocite{*}
\bibliography{aipsamp}
\section{SUPPLEMENTARY INFORMATION}
\clearpage
\appendix

\renewcommand{\thepage}{S\arabic{page}}  
\setcounter{page}{1}

\renewcommand{\thesection}{S\arabic{section}} 
\setcounter{section}{1}

\renewcommand{\thefigure}{S\arabic{figure}}   
\setcounter{figure}{0}

\renewcommand{\thetable}{S\arabic{table}}     
\setcounter{table}{1}

\begin{figure*}
\centering
\includegraphics[width=1.0\textwidth]{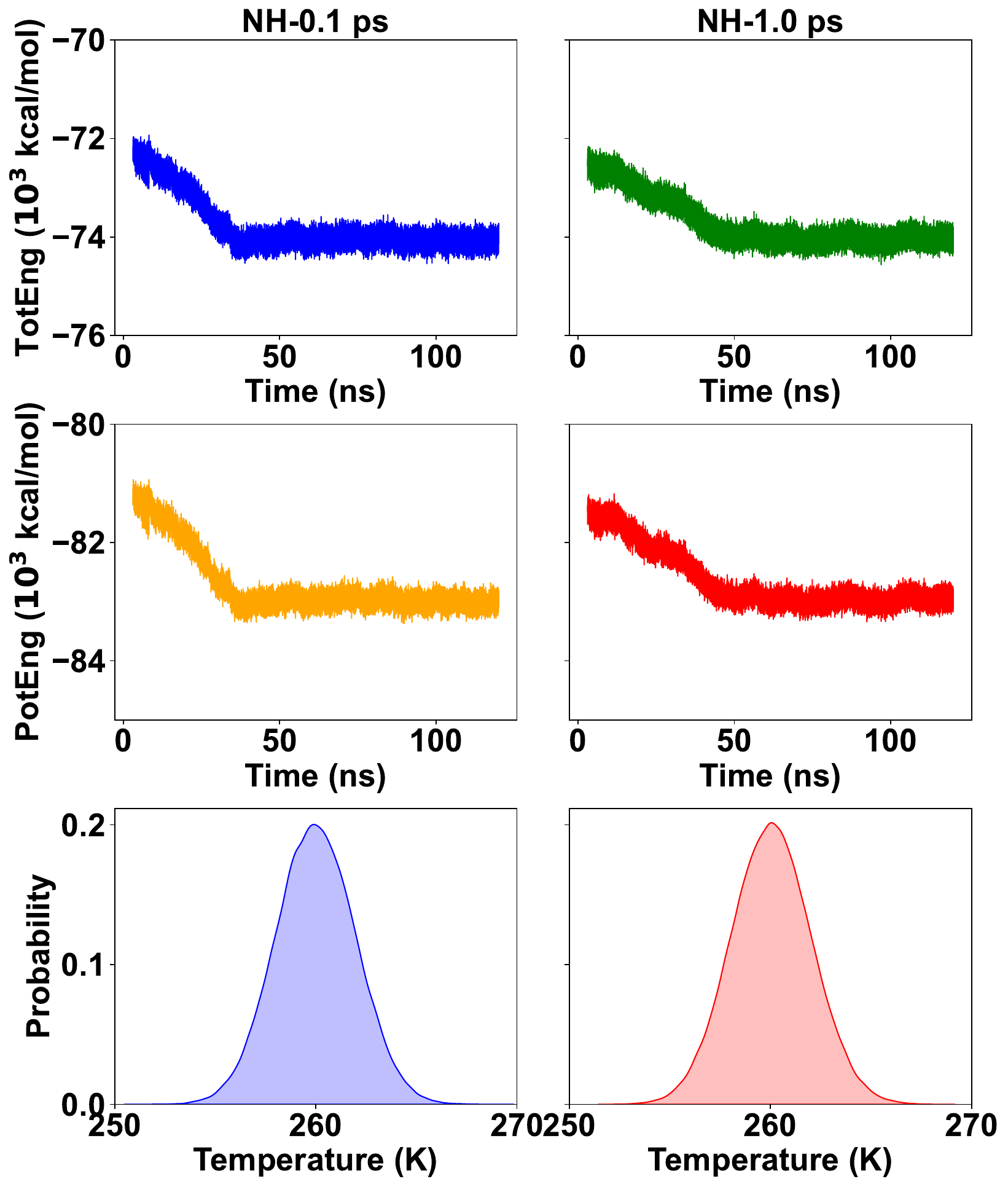}
\caption{The effects of different time constants (0.1 ps (left) and 1.0 ps (right)) for the Nos\'e-Hoover thermostat during the crystallization process at 260 K on the total energy (top), potential energy (middle), and temperature distribution (bottom).}
\label{nose-0.1ps-1ps}
\end{figure*}

\begin{figure*}
\centering
\includegraphics[width=1.0\textwidth]{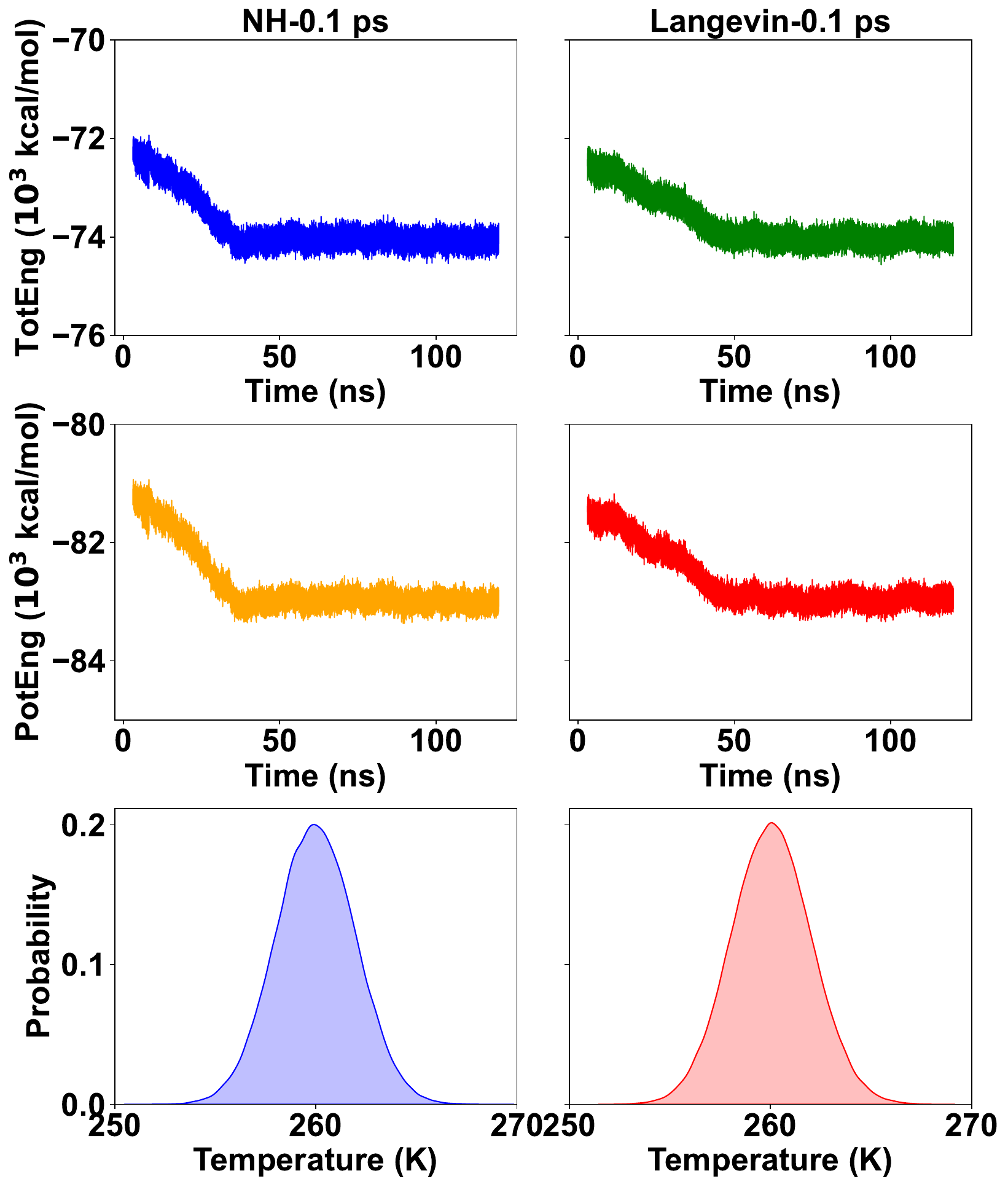}
\caption{The effects of the Nos\'e-Hoover (left) and Langevin (right) thermostat with same time constant of 0.1 ps during the crystallization process at 260 K on the total energy (top), potential energy (middle), and temperature distribution (bottom).}
\label{nose-0.1ps-langevin-0.1ps}
\end{figure*}

\begin{figure*}
\centering
\includegraphics[width=1.0\textwidth]{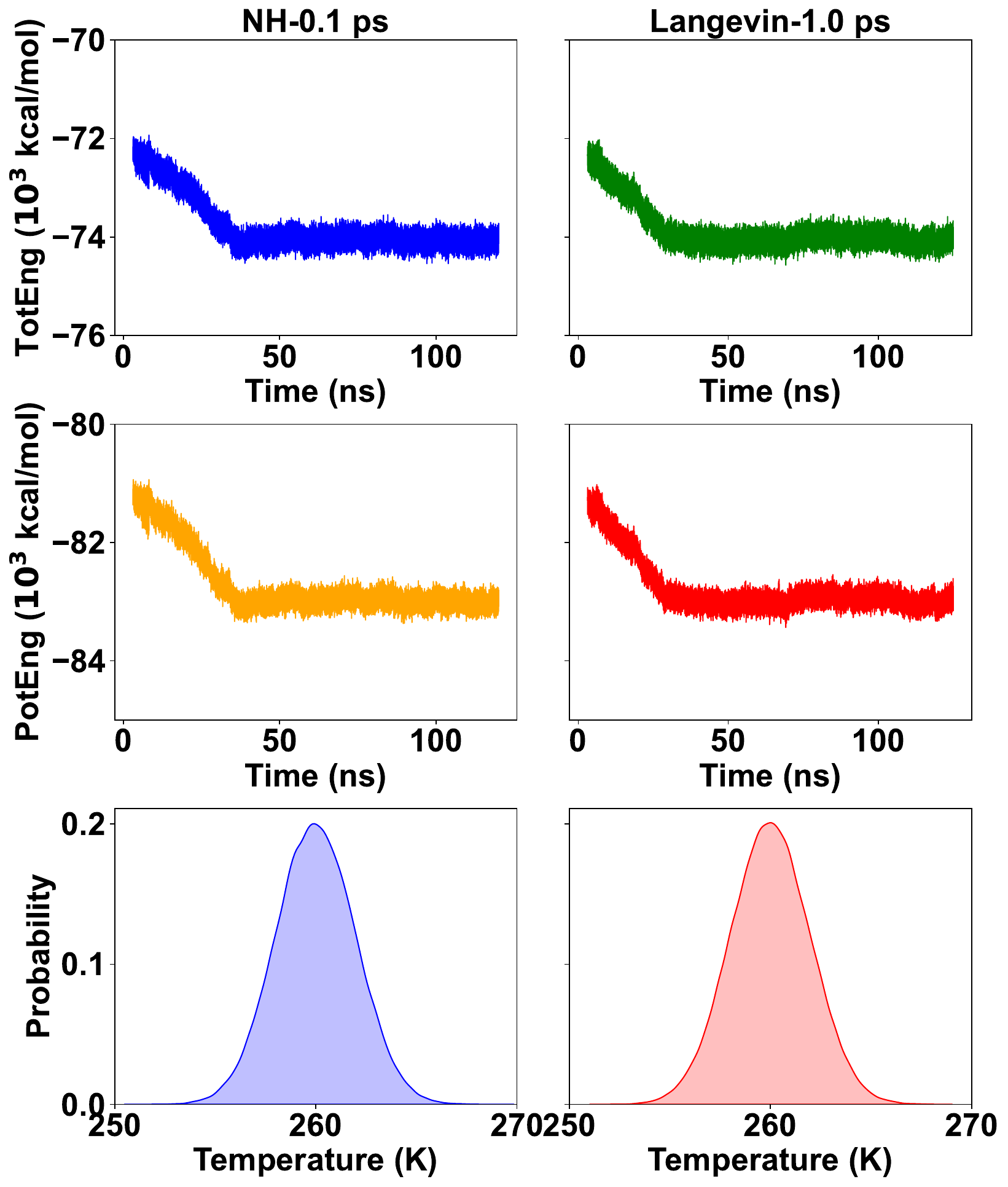}
\caption{The effects of different time constants for the Nos\'e-Hoover (left, 0.1 ps) and Langevin thermostat (right, 1.0 ps) during the crystallization process at 260 K on the total energy (top), potential energy (middle), and temperature distribution (bottom).}
\label{nose-0.1ps-langevin-1ps}
\end{figure*}

\begin{figure*}
\centering
\includegraphics[width=1.0\textwidth]{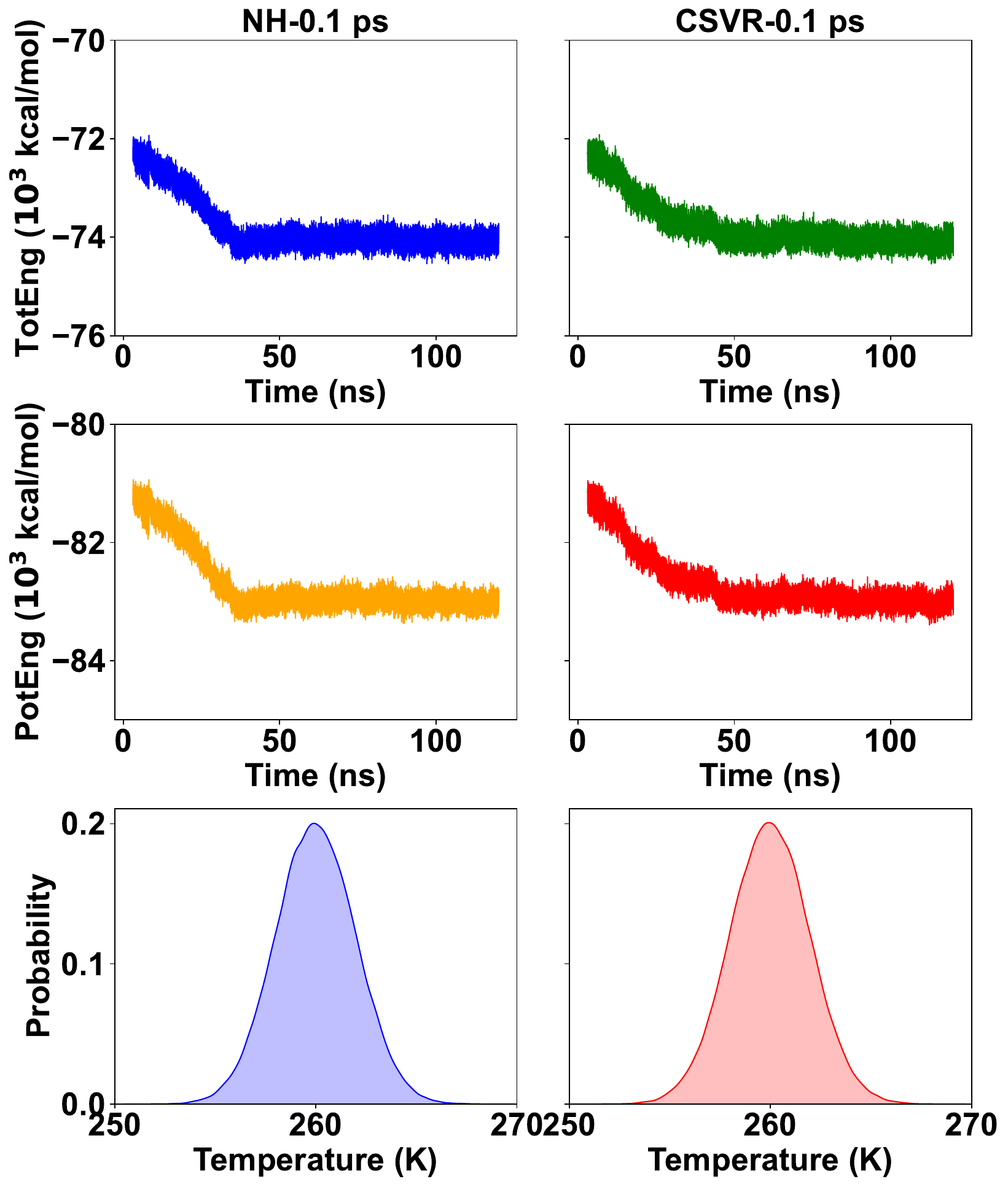}
\caption{The effects of the Nos\'e-Hoover (left) and CSVR (right) thermostat with same time constant of 0.1 ps during the crystallization process at 260 K on the total energy (top), potential energy (middle), and temperature distribution (bottom).}
\label{nose-0.1ps-csvr-0.1ps}
\end{figure*}

\begin{figure*}
\centering
\includegraphics[width=1.0\textwidth]{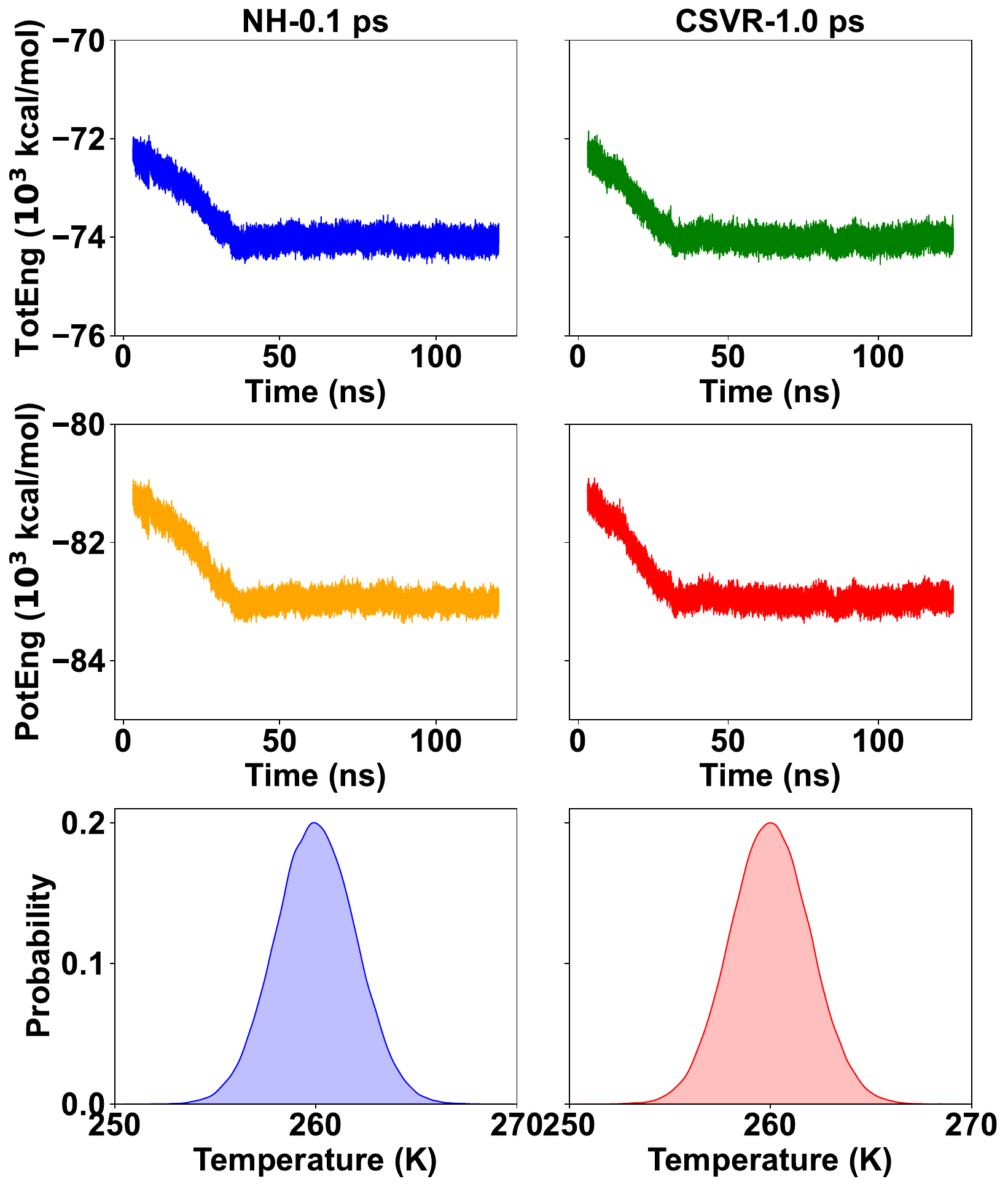}
\caption{The effects of different time constants for the Nos\'e-Hoover (left, 0.1 ps) and CSVR (right, 1.0 ps) thermostat during the crystallization process at 260 K on the total energy (top), potential energy (middle), and the temperature (bottom).}
\label{nose-0.1ps-csvr-1ps}
\end{figure*}

\begin{figure*}
\centering
\includegraphics[width=1.0\textwidth]{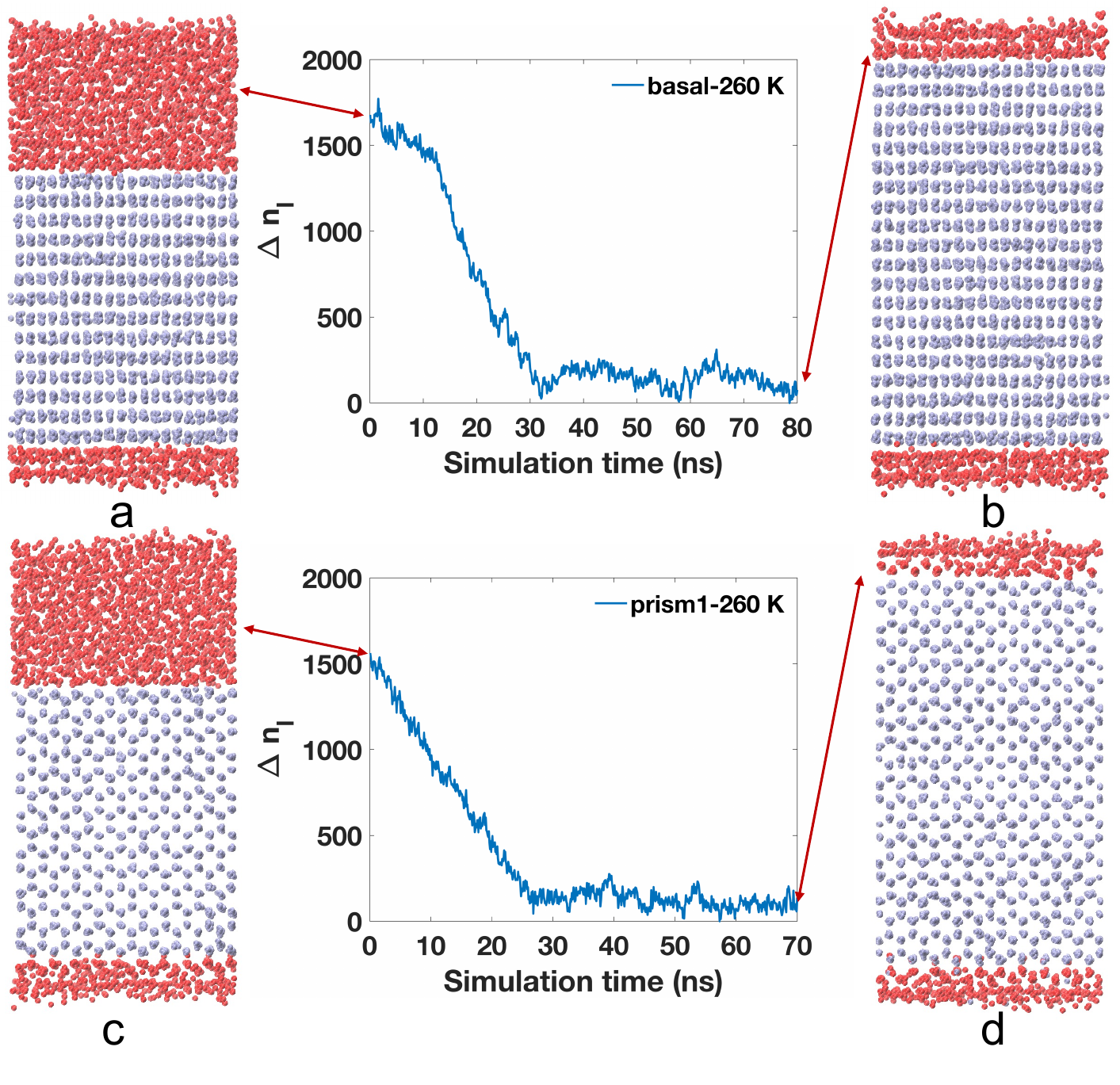}
\caption{(a) The initial and (b) the final structures in an 80 ns recrystallization simulation at 260 K for the basal surface. (c) The initial and (d) the final structures in a 70 ns recrystallization simulation at 260 K for the prism1 surface. The liquid-like molecules are coloured in red, and the ice-like molecules are in blue. Only oxygen atoms are shown.}
\label{melting_growth}
\end{figure*}

\begin{figure*}[htbp]
\centering
\includegraphics[width=1.1\textwidth]{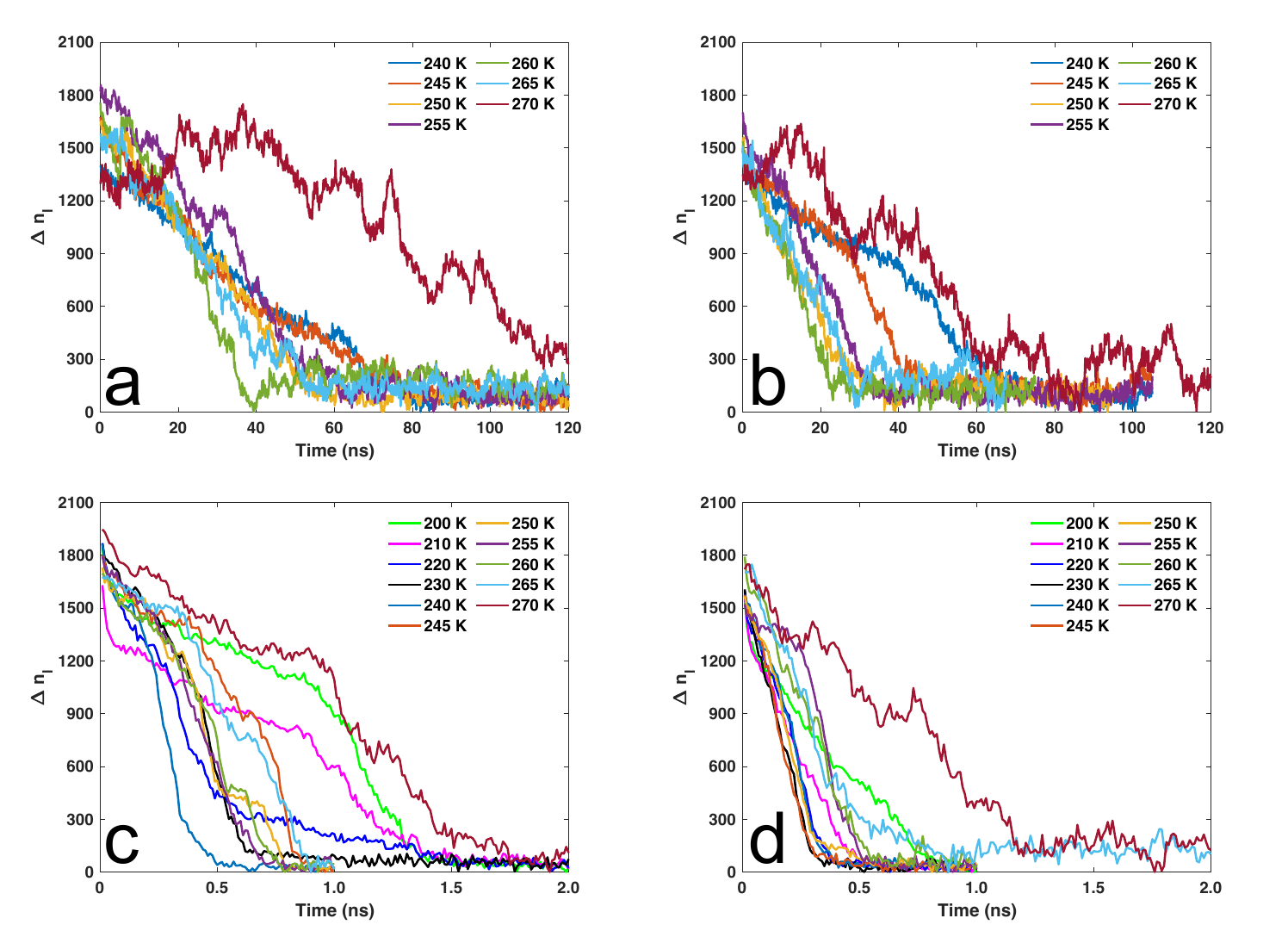}
\caption{Time-evolution of the excess number of liquid-like molecules $\Delta n_l$ during the TIP4P/Ice crystallization simulations for one replica at each simulated temperature for the basal surface (a) and for the prism1 surface (b). The time-evolution of $\Delta n_l$ during the mW crystallization simulations for one replica at each simulated temperature for the basal surface (c) and for the prism1 surface (d).}
\label{envolution_SI}
\end{figure*}

\begin{figure*}[htbp!]
\centering
\includegraphics[width=0.495\textwidth]{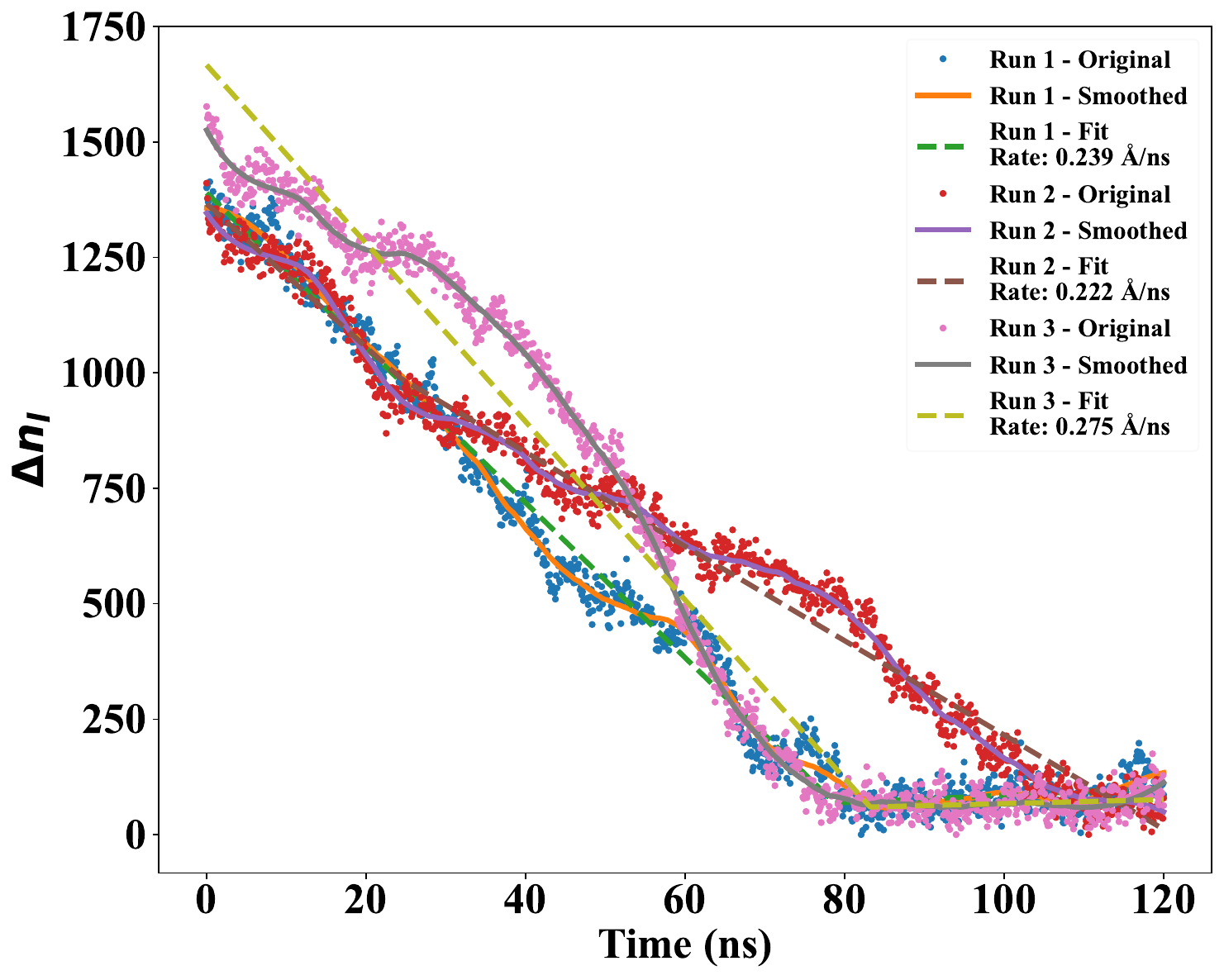}
\includegraphics[width=0.495\textwidth]{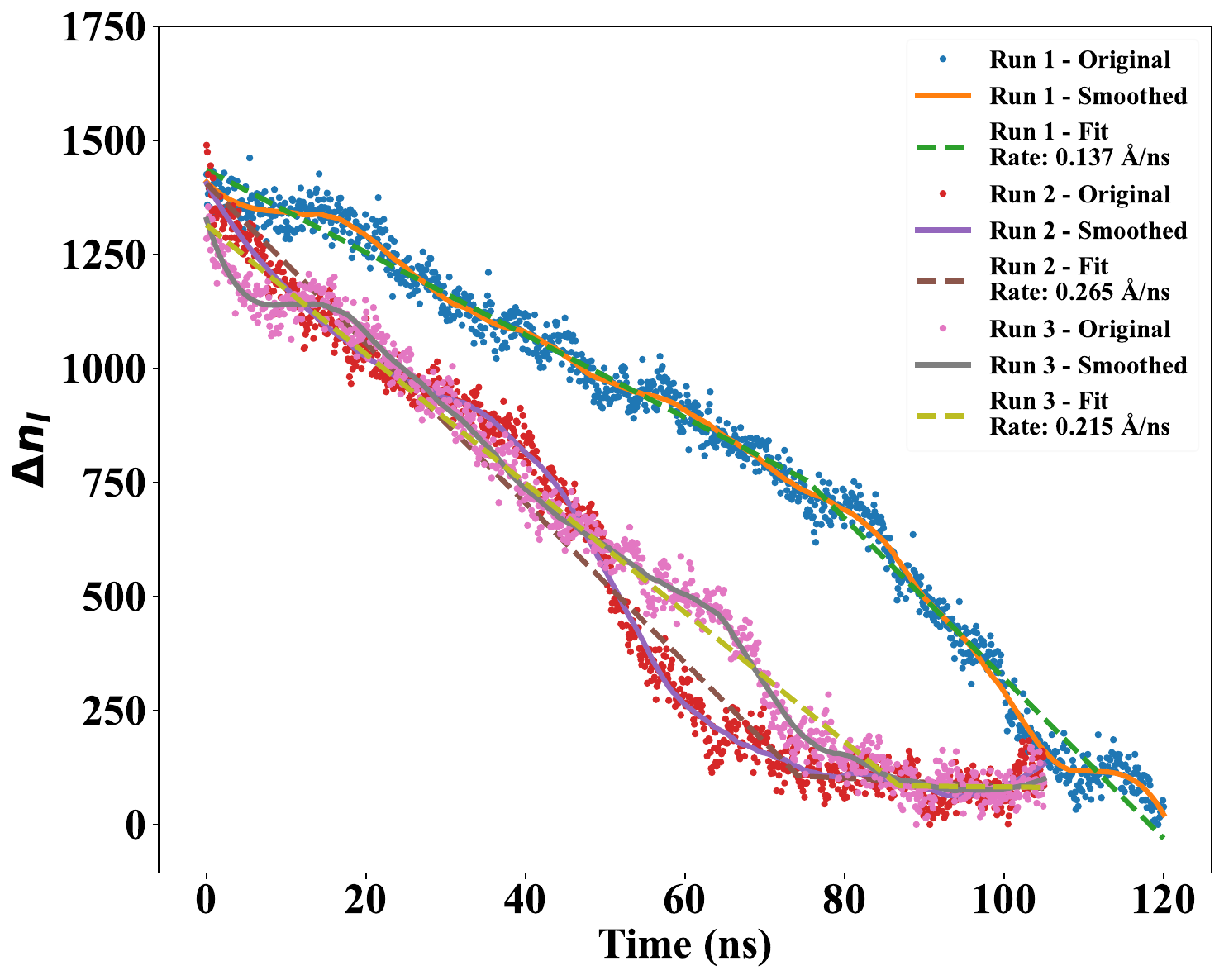}
\caption{Time-evolution of $\Delta n_l$ during the TIP4P/Ice crystallization simulations for the replicas of the ice Ih basal (left panel) and prism1 (right panel) surfaces at 240 K. }
\label{240K}
\end{figure*}

\begin{figure*}[htbp!]
\centering
\includegraphics[width=0.495\textwidth]{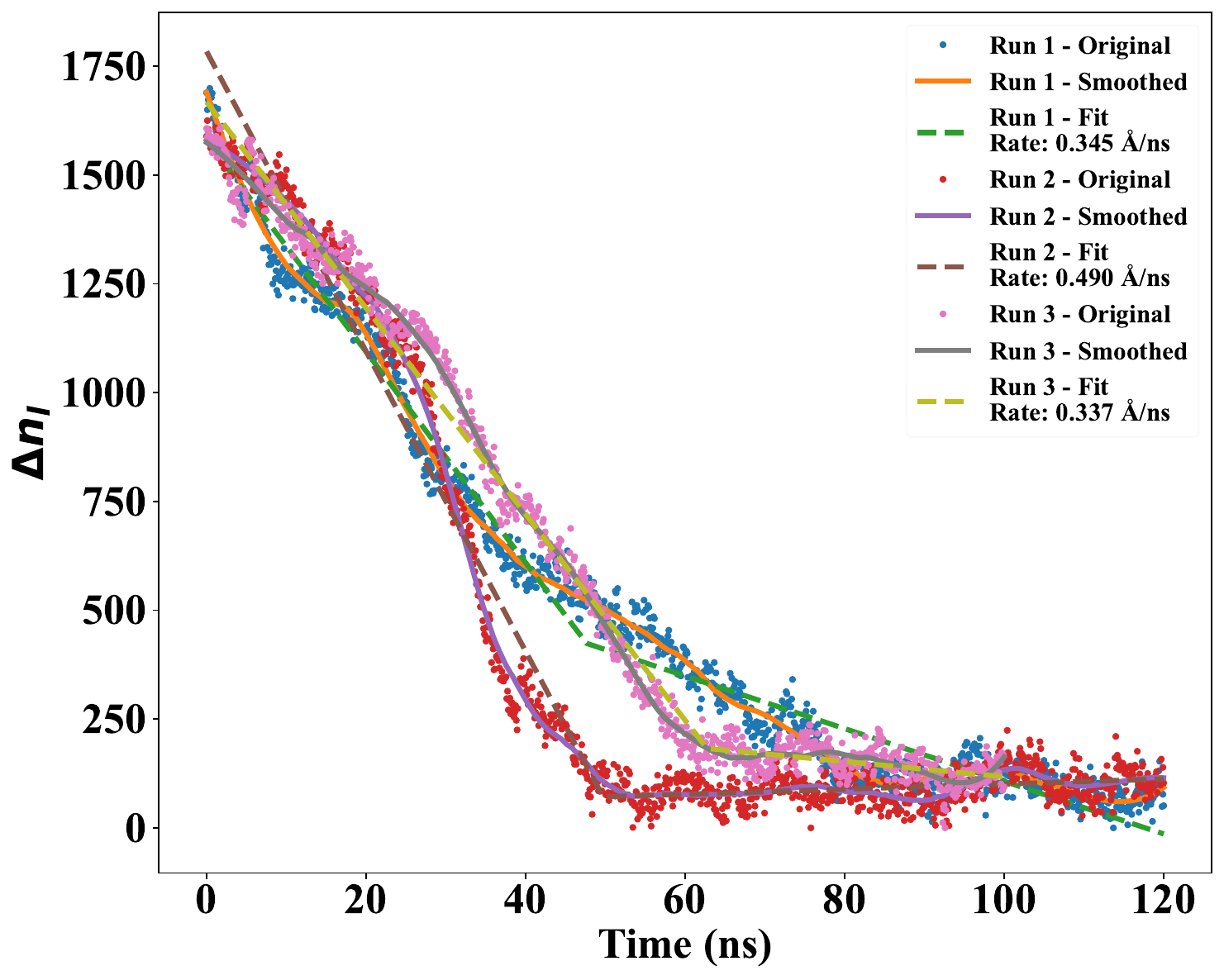}
\includegraphics[width=0.495\textwidth]{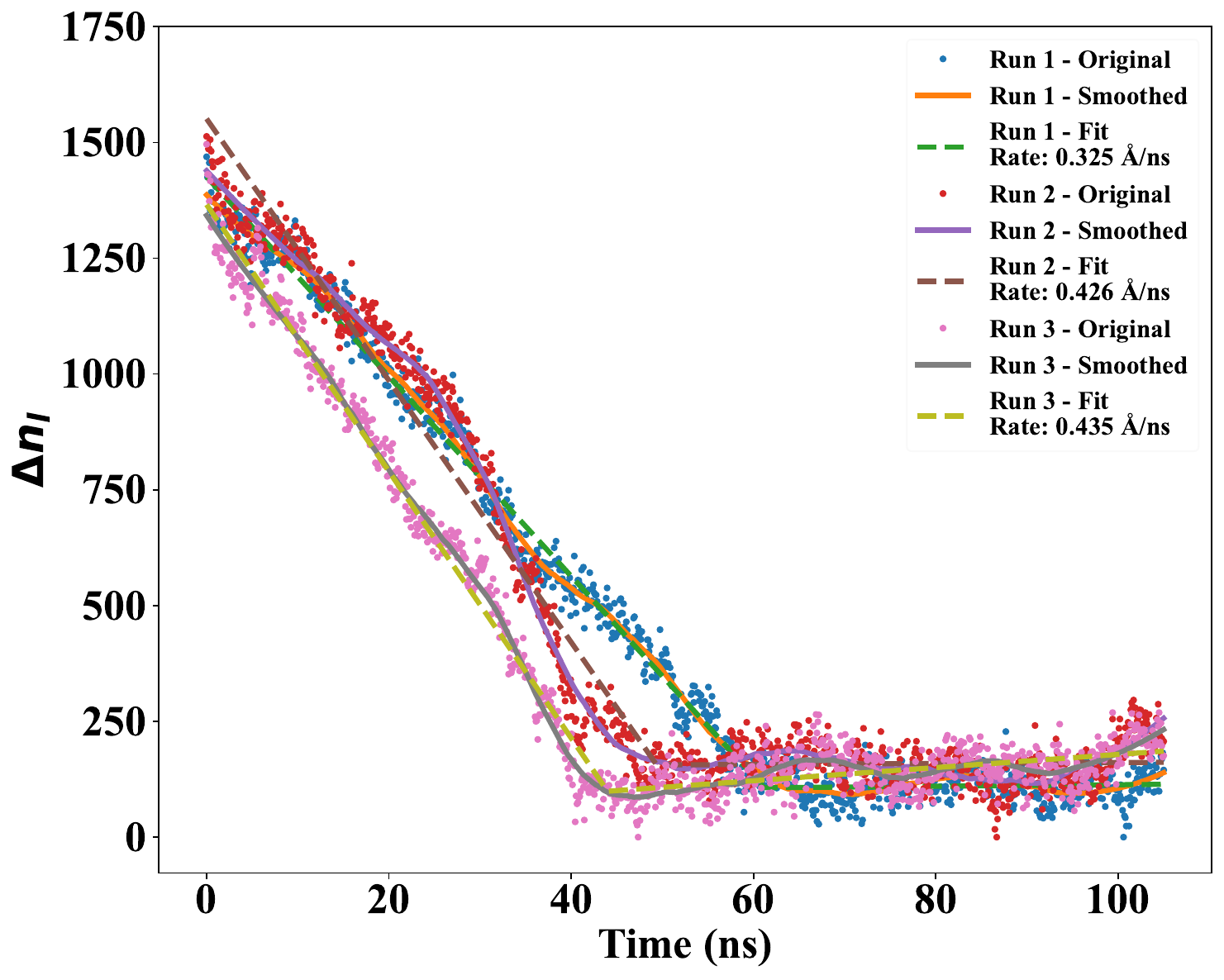}
\caption{Time-evolution of $\Delta n_l$ during the TIP4P/Ice crystallization simulations for the replicas of the ice Ih basal (left panel) and prism1 (right panel) surfaces at 245 K. }
\label{245k}
\end{figure*}

\begin{figure*}[htbp!]
\centering
\includegraphics[width=0.495\textwidth]{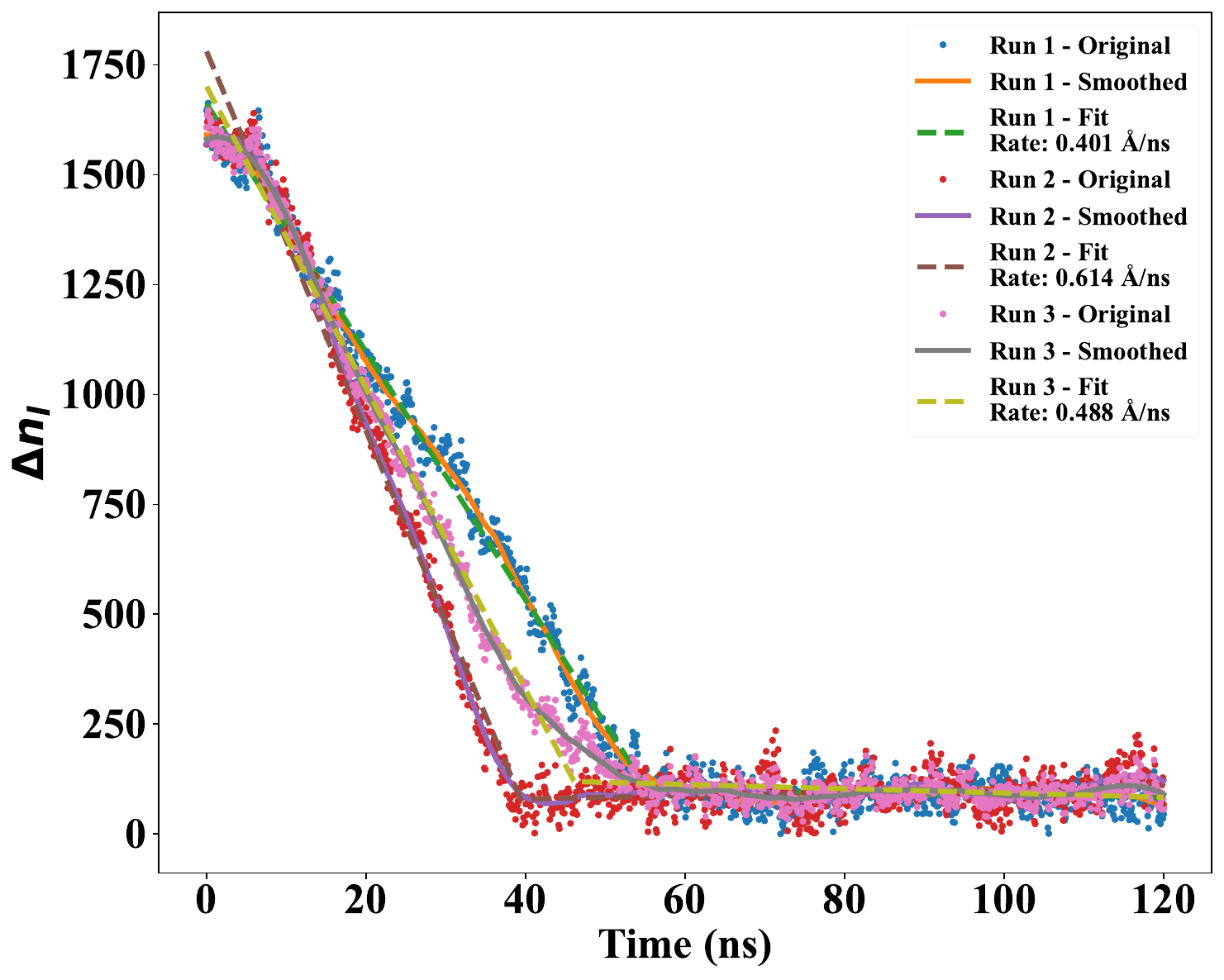}
\includegraphics[width=0.495\textwidth]{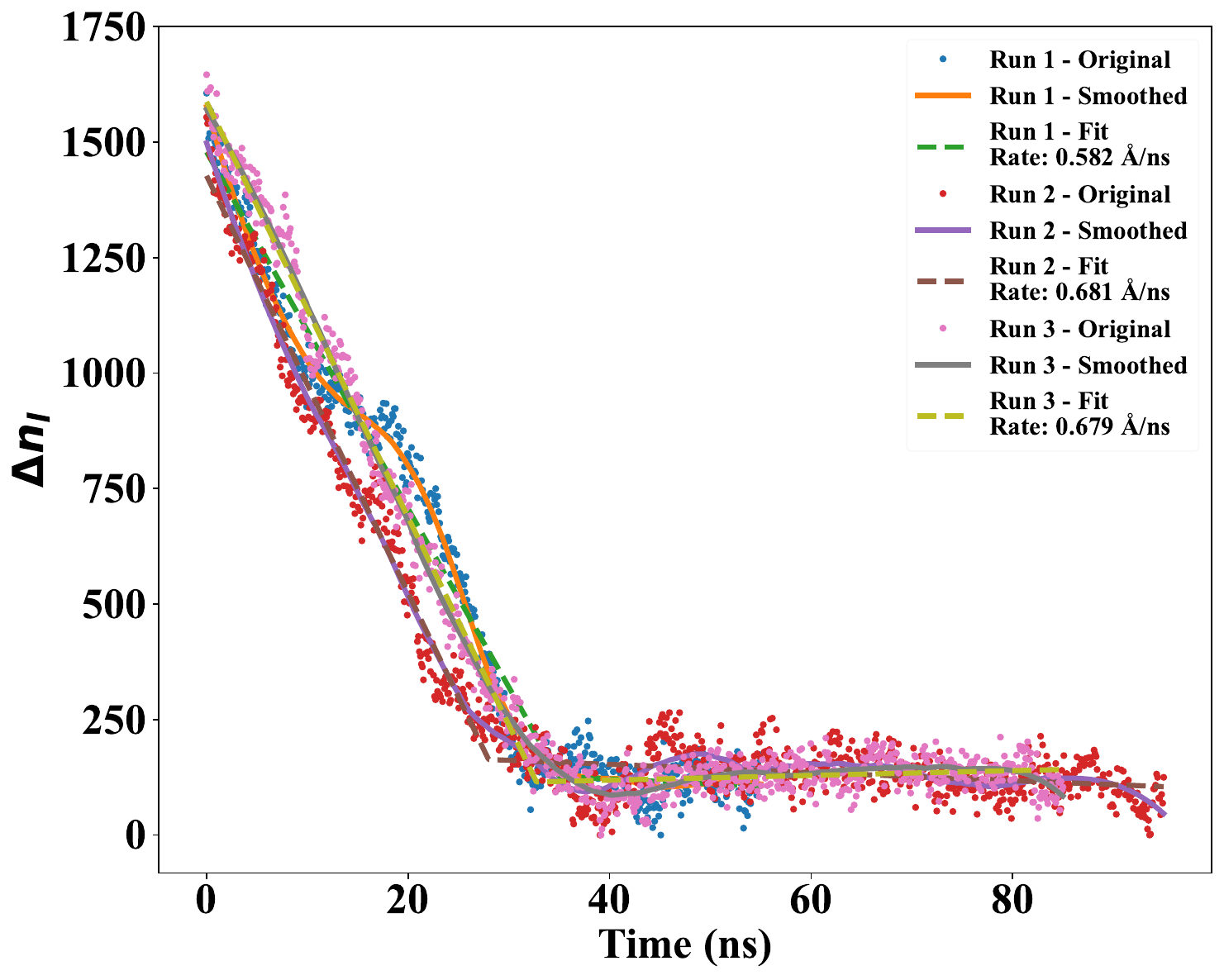}
\caption{Time-evolution of $\Delta n_l$ during the TIP4P/Ice crystallization simulations for the replicas of the ice Ih basal (left panel) and prism1 (right panel) surfaces at 250 K. }
\label{250k}
\end{figure*}

\begin{figure*}[htbp!]
\centering
\includegraphics[width=0.495\textwidth]{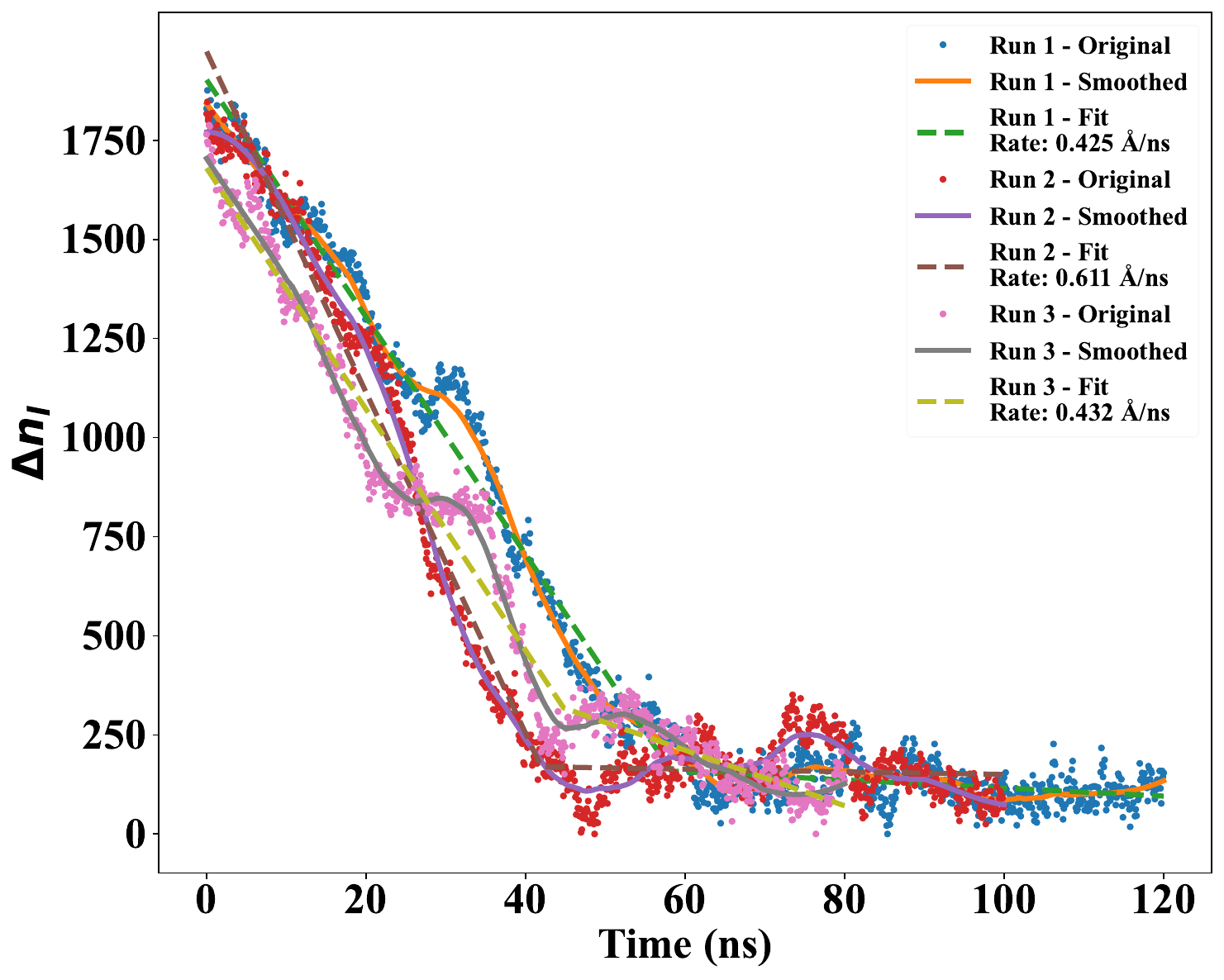}
\includegraphics[width=0.495\textwidth]{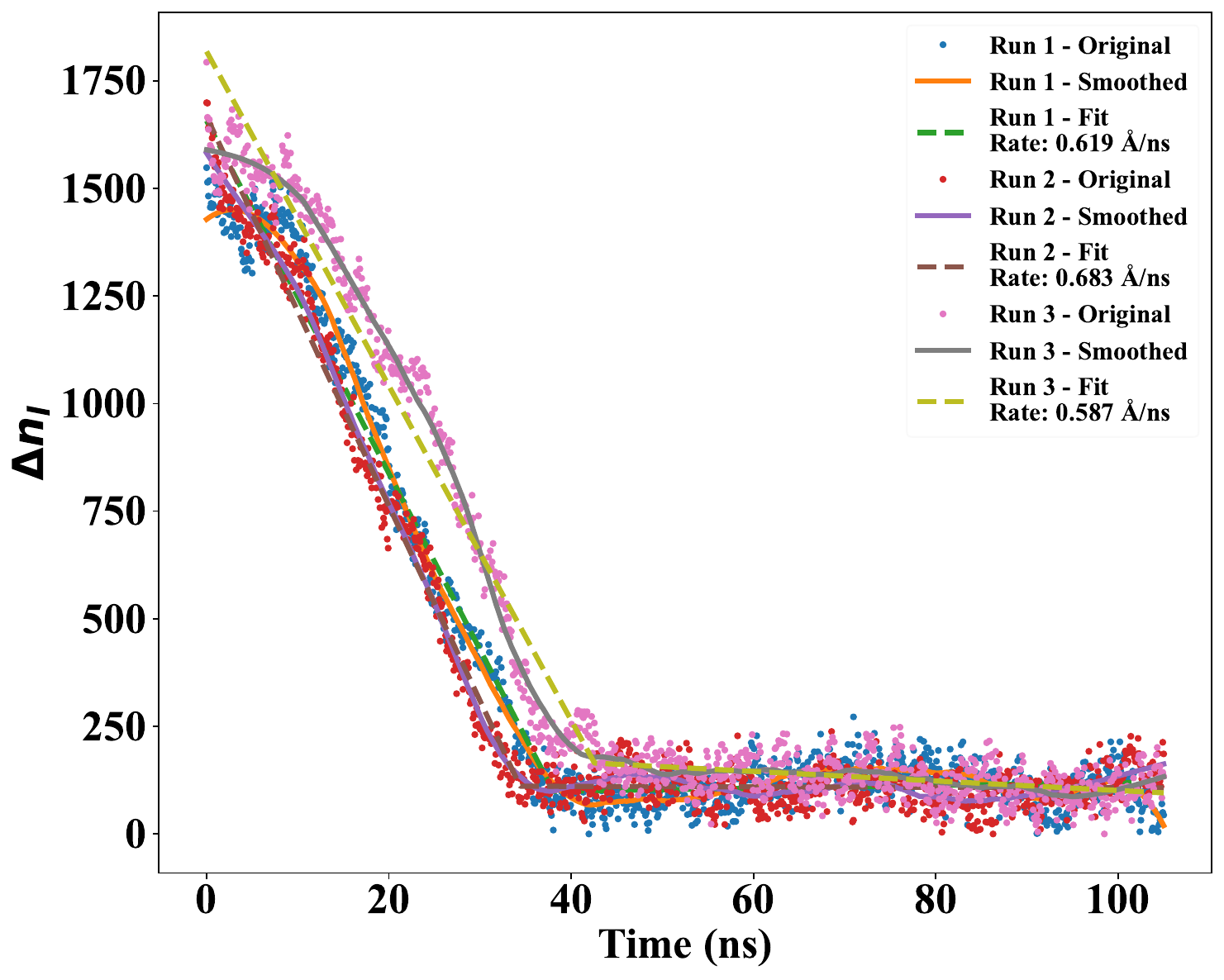}
\caption{Time-evolution of $\Delta n_l$ during the TIP4P/Ice crystallization simulations for the replicas of the ice Ih basal (left panel) and prism1 (right panel) surfaces at 255 K. }
\label{255k}
\end{figure*}

\begin{figure*}[htbp!]
\centering
\includegraphics[width=0.495\textwidth]{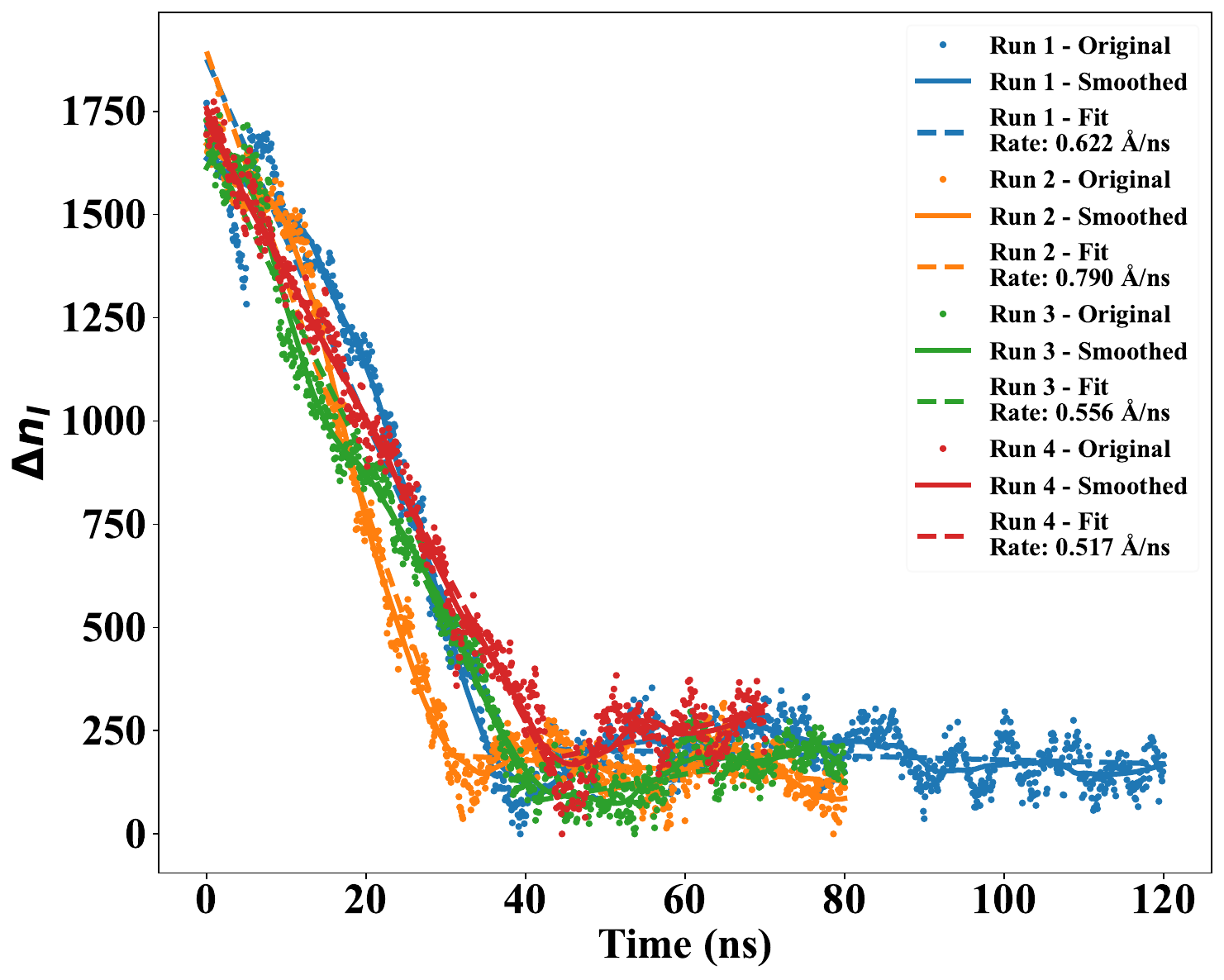}
\includegraphics[width=0.495\textwidth]{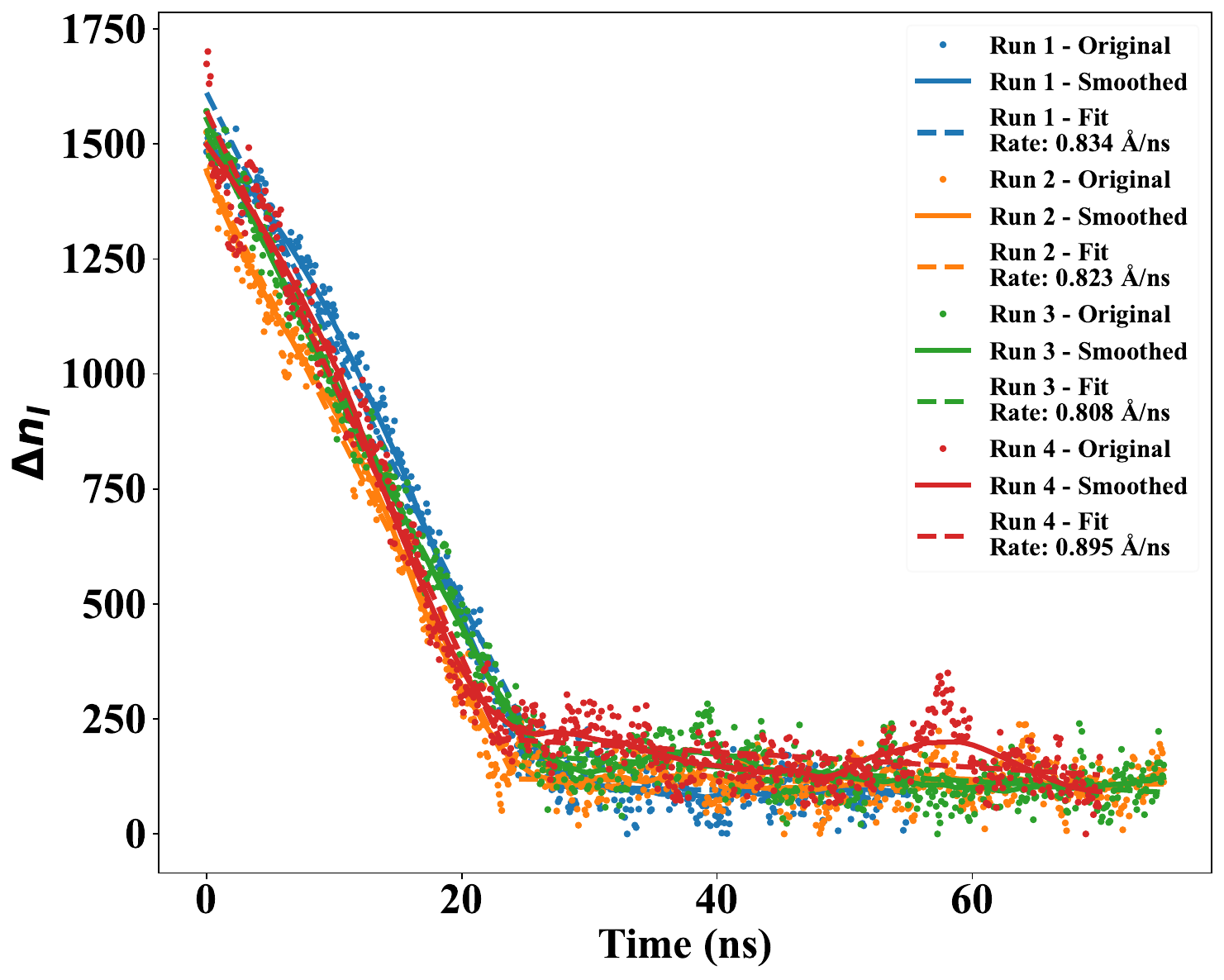}
\caption{Time-evolution of $\Delta n_l$ during the TIP4P/Ice crystallization simulations for the replicas of the ice Ih basal (left panel) and prism1 (right panel) surfaces at 260 K. }
\label{260k}
\end{figure*}

\begin{figure*}[htbp!]
\centering
\includegraphics[width=0.495\textwidth]{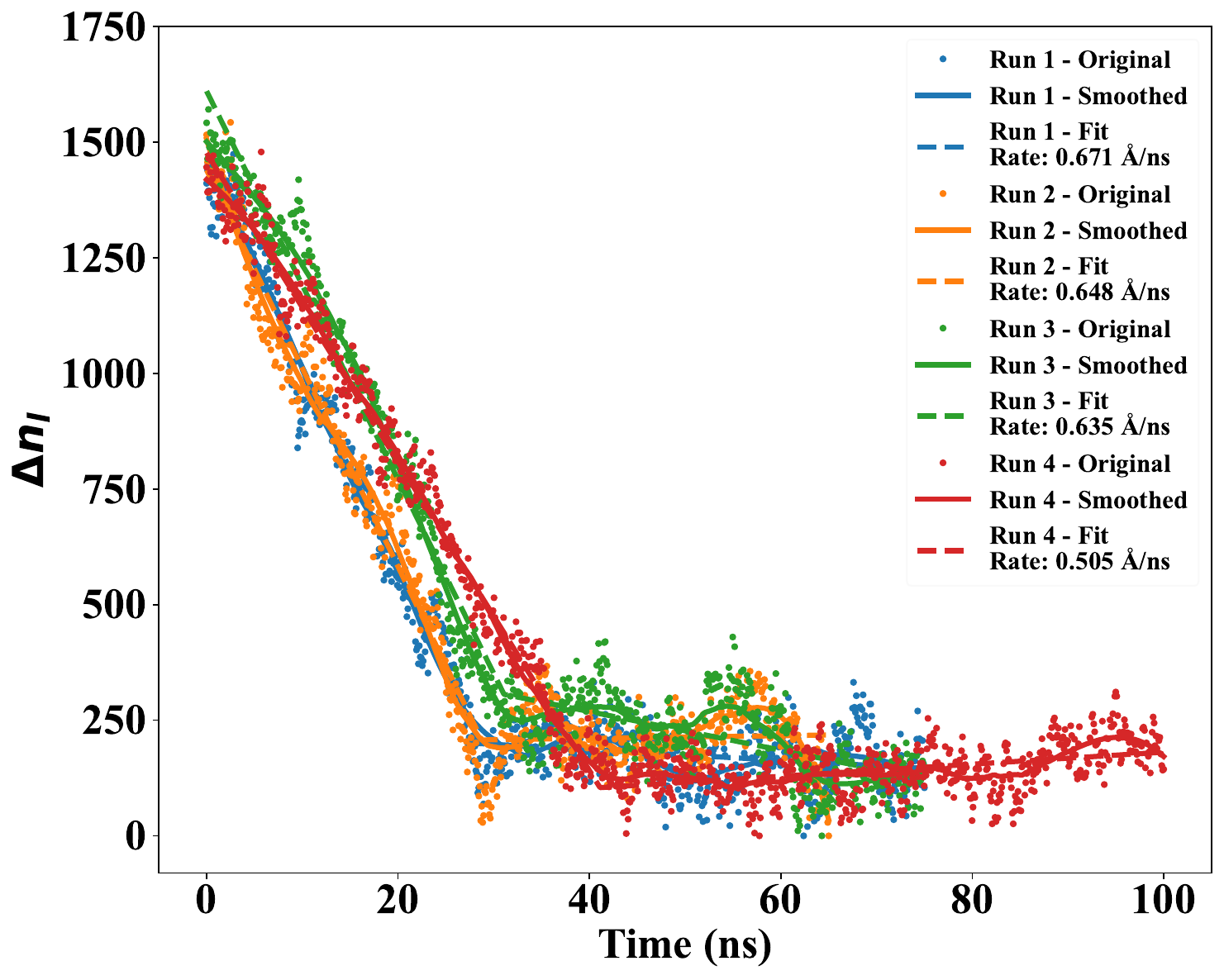}
\includegraphics[width=0.495\textwidth]{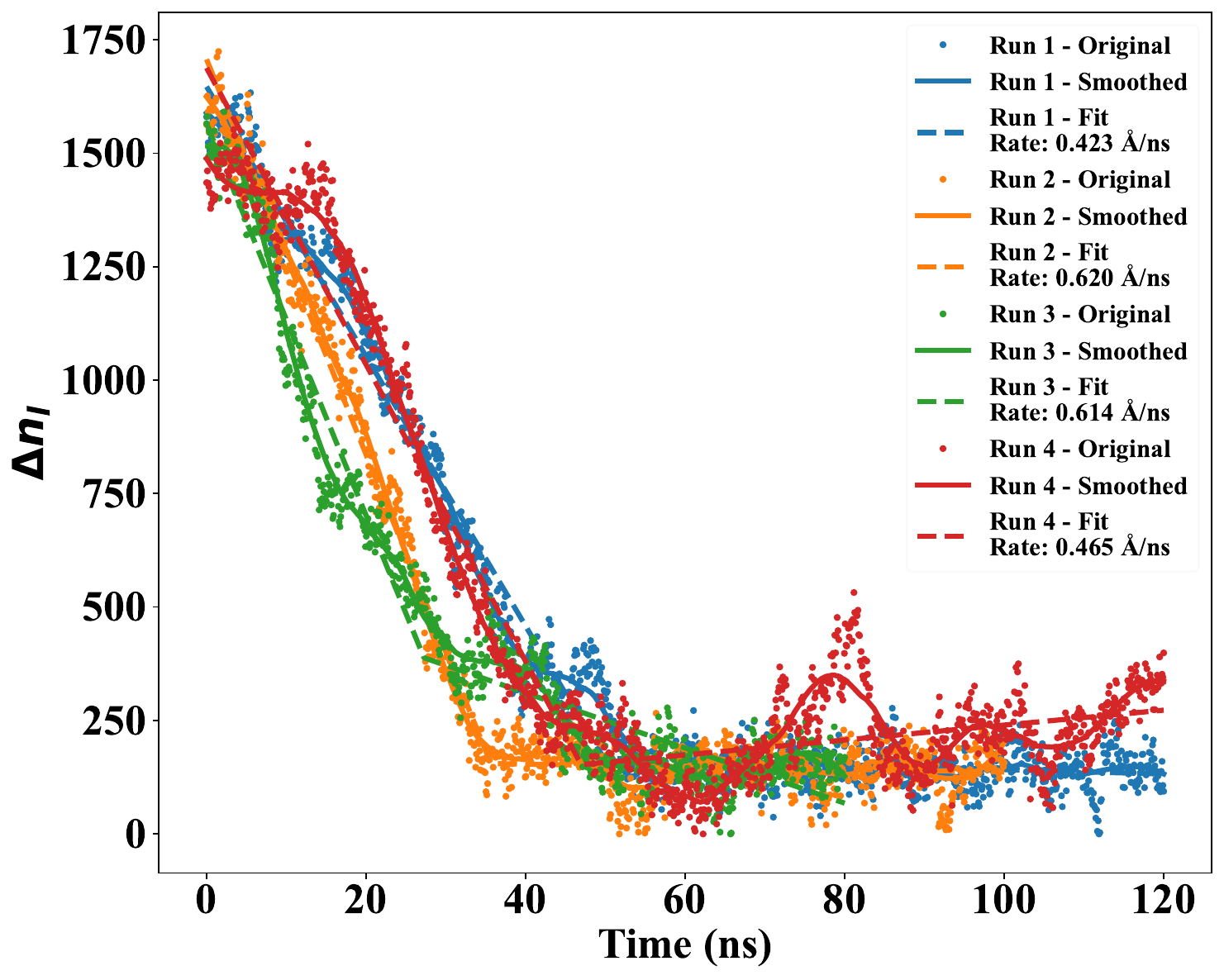}
\caption{Time-evolution of $\Delta n_l$ during the TIP4P/Ice crystallization simulations for the replicas of the ice Ih basal (left panel) and prism1 (right panel) surfaces at 265 K. }
\label{265k}
\end{figure*}

\begin{figure*}[htbp!]
\centering
\includegraphics[width=0.495\textwidth]{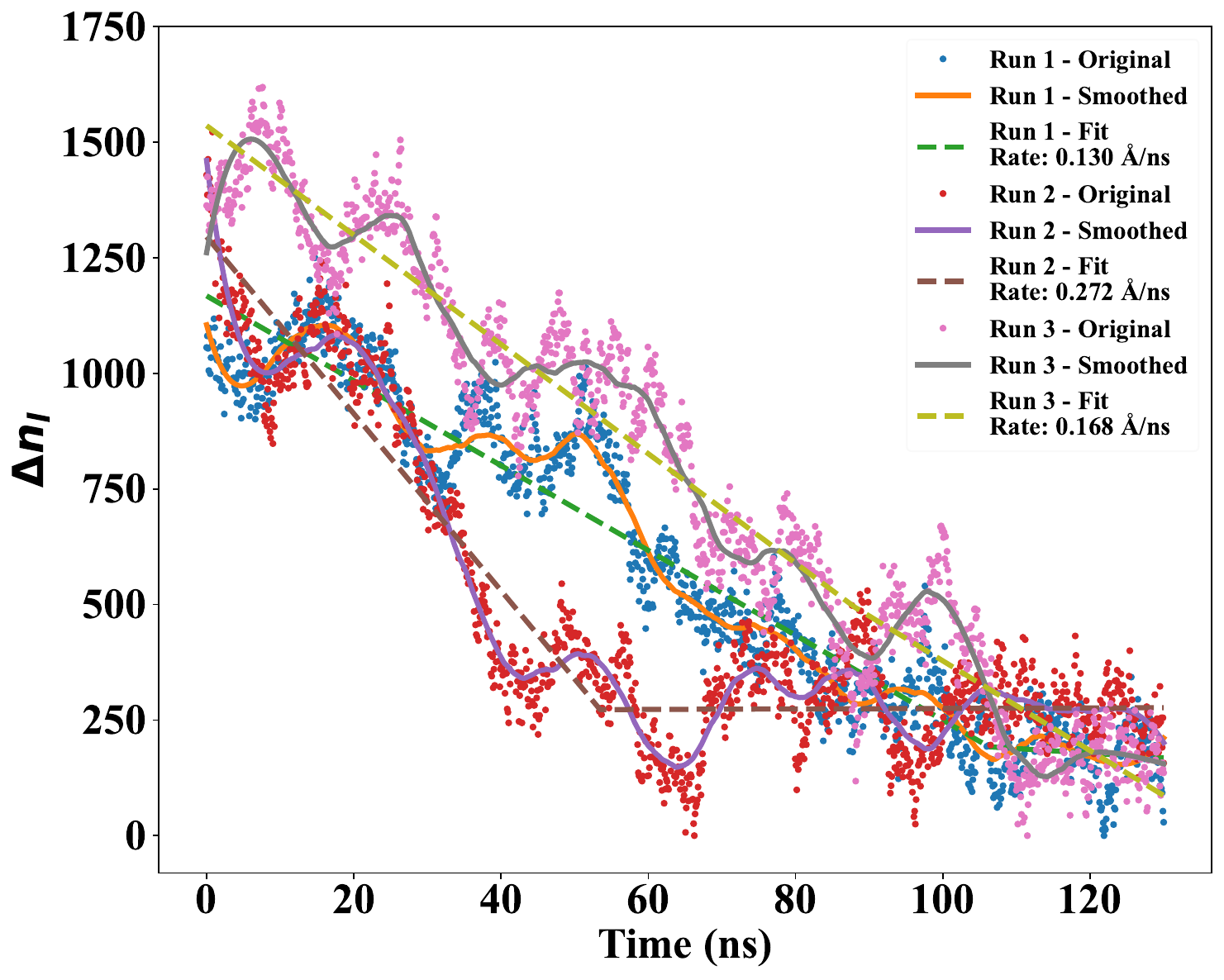}
\includegraphics[width=0.495\textwidth]{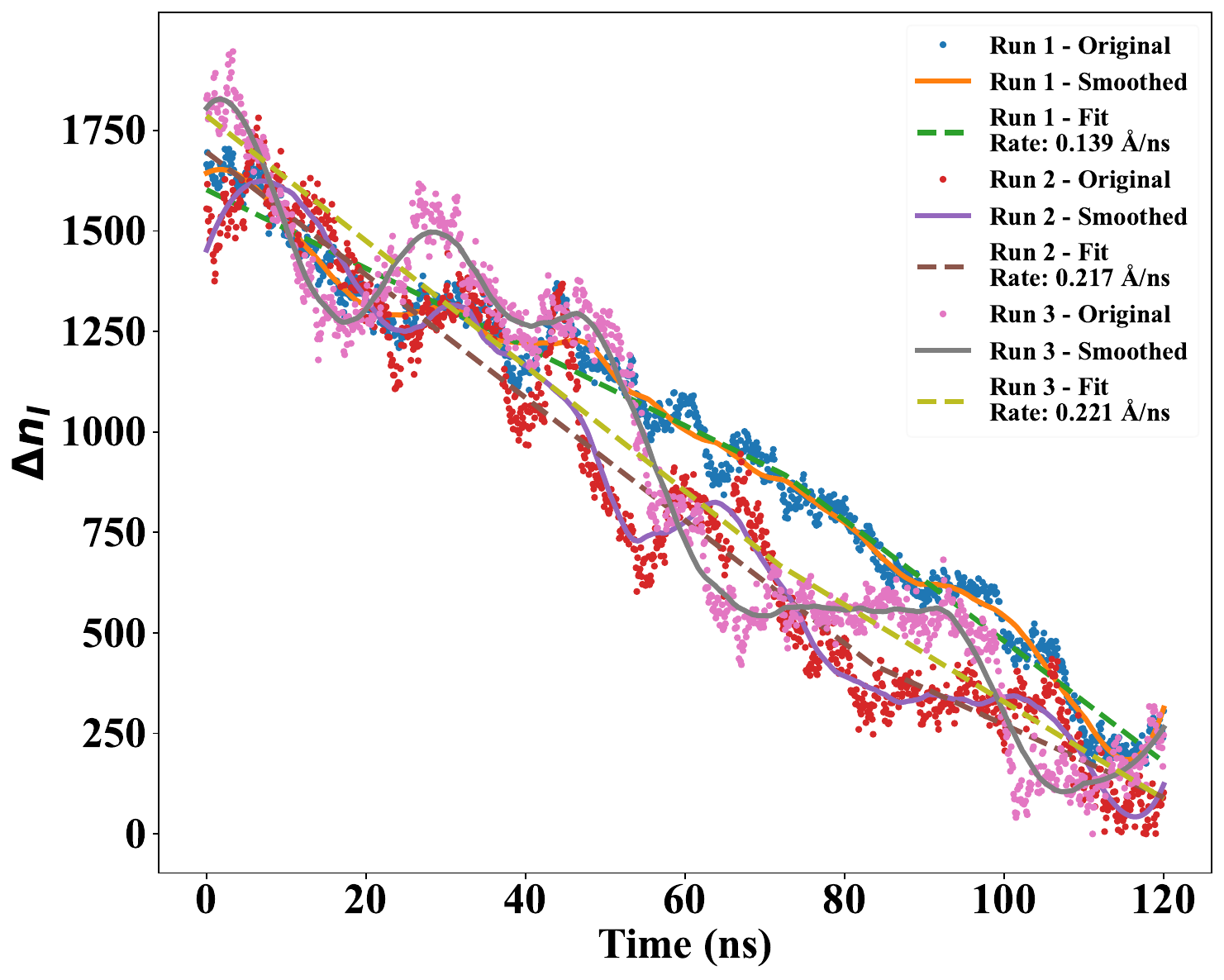}
\caption{Time-evolution of $\Delta n_l$ during the TIP4P/Ice crystallization simulations for the replicas of the ice Ih basal (left panel) and prism1 (right panel) surfaces at 270 K.} 
\label{270k}
\end{figure*}

\begin{figure*}[htbp!]
\centering
  \includegraphics[width=0.85\textwidth]{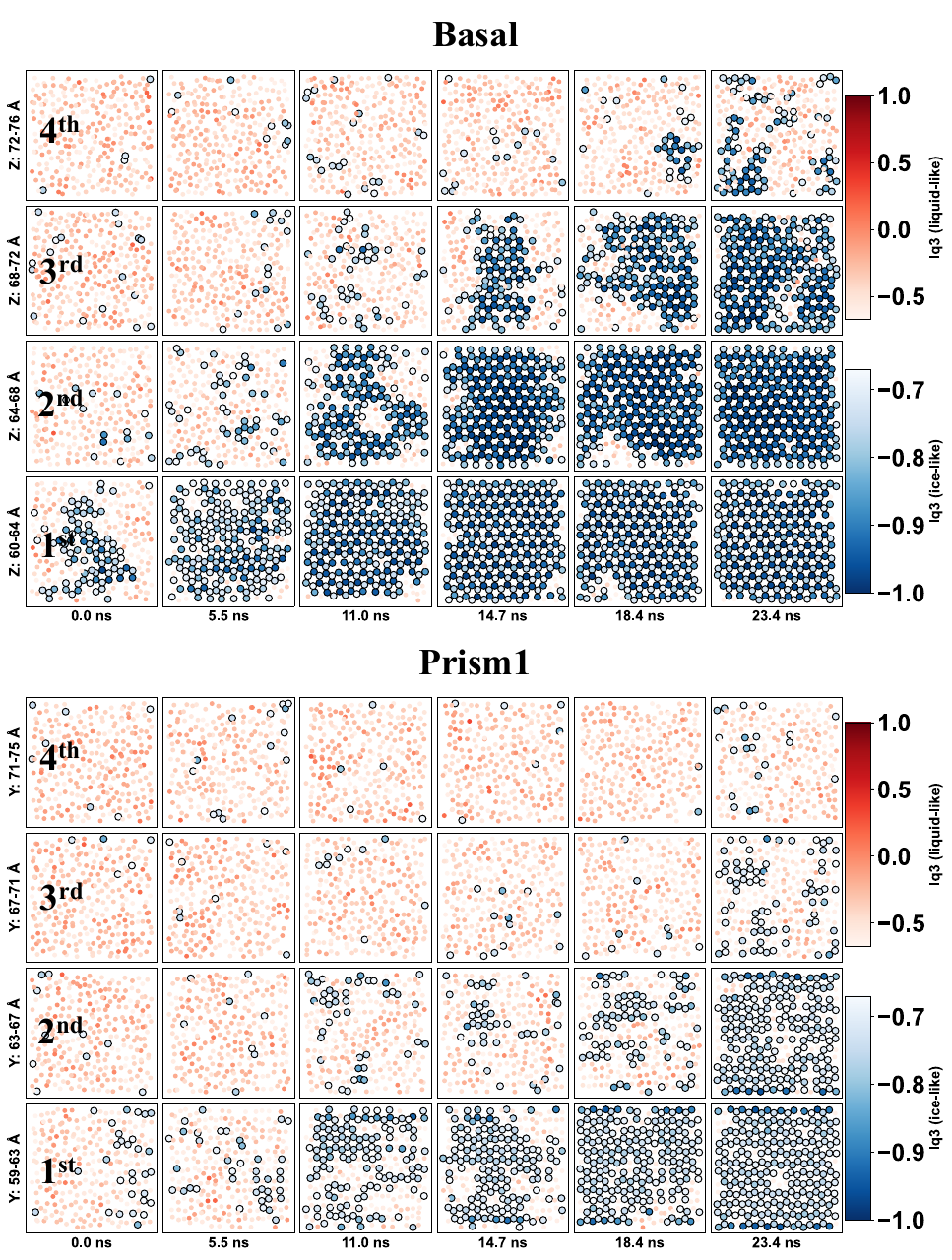} 
  \caption{Time-evolution of the formation of the four ice bilayers above the interface where the melted area meets the ice region for the basal (top panel) and prism1 (bottom panel) surfaces during the TIP4P/Ice crystallization process at 265 K. The darker blue indicates that the molecules are in a more ordered state, i.e., ice-like molecules; the more reddish the dots are, the more disordered the molecules are, i.e., liquid-like molecules. The oxygen atoms are coloured according to the values of the order parameter $lq_3$.}
  \label{tip4p_kinetics-265K}
\end{figure*}

\end{document}